\documentclass[10pt,twocolumn,letterpaper]{article}

\usepackage[spanish,english]{babel}
\usepackage[utf8x]{inputenc}
\usepackage[T1]{fontenc}

\usepackage[a4paper,top=3cm,bottom=2cm,left=1.2cm,right=1.2cm,marginparwidth=1.75cm]{geometry}

\usepackage{amsmath}
\usepackage{graphicx}
\usepackage[colorinlistoftodos]{todonotes}
\usepackage[colorlinks=true, allcolors=blue]{hyperref}

\usepackage{graphicx}
\usepackage{amssymb}
\usepackage{pdflscape}
\usepackage{multirow} 
\usepackage{rotating}
\usepackage{url} 
\usepackage{longtable}

\usepackage{fancyvrb}
\usepackage{fvextra}

\usepackage{array}
\usepackage{tabularx}
\usepackage{multirow}
\usepackage{xcolor}

\usepackage{hyperref}
\usepackage{makecell}

\newcommand{\cualpha}{cu\textrm{-}\alpha}
\newcommand{\Cualpha}{Cu\textrm{-}\alpha}

\title{Applying Inter-rater Reliability and Agreement in Grounded Theory Studies in Software Engineering}

\usepackage{authblk}
\author[1,*]{Jessica D\'iaz}
\author[1]{Jorge P\'erez}
\author[1]{Carolina Gallardo}
\author[2,3]{\'Angel Gonz\'alez-Prieto}

\affil[1]{Universidad Polit\'ecnica de Madrid. Departamento de Sistemas Inform\'aticos. ETSI Sistemas Inform\'aticos, C.\ Alan Turing s/n (Carretera de Valencia Km 7), 28031 Madrid, Spain.}
\affil[2]{Universidad Complutense de Madrid. Departamento de \'Algebra, Geometr\'ia y Topolog\'ia,\newline{}Facultad de Ciencias Matem\'aticas, Plaza Ciencias 3, 28040 Madrid, Spain.}
\affil[3]{Instituto de Ciencias Matem\'aticas (CSIC-UAM-UCM-UC3M), \newline{}C.\ Nicol\'as Cabrera 13-15, 28049 Madrid, Spain.}

\affil[*]{Corresponding author: yesica.diaz@upm.es}

\date{}

\begin{document}

\maketitle

\begin{abstract}
\parindent=0mm

\textit{Context}: In recent years, the qualitative research on empirical software engineering that applies Grounded Theory is increasing. Grounded Theory (GT) is a technique for developing theory inductively e iteratively from qualitative data based on theoretical sampling, coding, constant comparison, memoing, and saturation, as main characteristics. Large or controversial GT studies may involve multiple researchers in collaborative coding, which requires a kind of rigor and consensus that an individual coder does not. Although many qualitative researchers reject quantitative measures in favor of other qualitative criteria, many others are committed to measuring consensus through Inter-Rater Reliability (IRR) and/or Inter-Rater Agreement (IRA) techniques to develop a shared understanding of the phenomenon being studied. However, there are no specific guidelines about how and when to apply IRR/IRA during the iterative process of GT, so researchers have been using ad hoc methods for years.
\textit{Objective:} This paper presents a process for systematically applying IRR/IRA in GT studies that meets the iterative nature of this qualitative research method, which is supported by a previous systematic literature review on applying IRR/RA in GT studies in software engineering. This process allows researchers to incrementally generate a theory while ensuring consensus on the constructs that support it and, thus, improving the rigor of qualitative research.    
\textit{Method:} Meta-science guided us to analyze the issues and challenges of the GT method when various raters are involved in coding procedures and formalize a process to improve collaborative team science and consortia.
\textit{Results:} This process formalization helps researchers to apply IRR/IRA to GT studies 
when various raters are involved in coding. Measuring consensus among raters promotes communicability, transparency, reflexivity, replicability, and trustworthiness of the research.
\textit{Conclusions:} The application of this process to a GT study seems to support its feasibility. In the absence of further confirmation, this would represent the first step of a de facto standard to be applied to those GT studies that require IRR/IRA.
\newline{}{\bf Keywords}: Grounded Theory, Inter-rater Reliability, Inter-rater Agreement.
\end{abstract}

\section{Introduction}

Qualitative data collection and analysis techniques are increasingly used in software engineering research \cite{STOL:2016,wohlin:2012,SALLEH:2018}. Among the most popular are the various flavours of Grounded Theory \cite{glaser:1967,strauss:1990,charmaz:2014}. \textit{Grounded Theory} (GT) refers to a family of predominately qualitative research methods for inductively generating theory based on rounds of interleaved data collection and analysis. Hence, GT is particularly well suited to exploring how software professionals collaborate and create software \cite{hoda2012,hoda2021}. 

GT involves one or more human analysts reading textual data (e.g.\ interview transcripts, documents, emails, discussion forum posts, field notes), labeling this data (\textit{coding}), recording their thoughts in notes called \textit{memos}, organizing the data and labels into categories and constantly comparing and reorganizing these categories until a mature or \textit{saturated} theory emerges. Many of the analytical techniques pioneered in the GT literature are now used in other qualitative methods including case studies, qualitative surveys, and critical discourse analysis because they help researchers examine and interpret qualitative data \cite{cruzes:2011,Saldana2012}. 

Collaborative coding, in which two or more researchers analyze qualitative data together or independently, is believed to improve research quality \cite{yin2017Case,dube2003rigor}. In fact, large or controversial GT studies often involve multiple researchers (collaborative team science and consortia), who frequently use \textit{collaborative coding} to develop a shared understanding of the phenomenon being studied \cite{Saldana2012}. Although there is no universal rule to determine whether two or more raters should be involved in coding procedures \cite{ralph2021}, collaborative coding or \textit{team coding} is increasingly popular \cite{erickson:1998,weston:2001,guest:2008}. 

However, team coding requires a kind of rigor that individual coding does not \cite{weston:2001} as well as consensus among raters \cite{campbell:2013,oconnor:2020}. When various raters are involved in the coding process, disagreements among raters may emerge. These disagreements may be resolved by discussion, voting, or an additional coder who acts as a tiebreaker. Nevertheless, when disagreements arise, researchers (and reviewers) may question the validity of the coding process.

Meanwhile, researchers often assess consensus among coders using a measure of Inter-Rater Reliability (IRR) or Inter-Rater Agreement (IRA) \cite{armstrong:1997,weston:2001,campbell:2013,MacPhail:2016,mcdonald:2019,oconnor:2020}. IRA is the extent to which different raters assign the same precise value for each item being rated, whereas IRR is the extent to which raters can consistently distinguish between different items on a measurement scale \cite{gisev2013interrater}. Briefly, IRA measures agreement whereas IRR measures consistency; raters may have high consistency but low (or even no) agreement.

However, deciding whether to use quantitative measures like IRR/IRA in qualitative research has little consensus with a lot of researchers in favor and in against \cite{mcdonald:2019,oconnor:2020}.  Additionally, while there are general guidelines for applying IRR/IRA \cite{MacPhail:2016,oconnor:2020}, there are no guidelines specific to grounded theory beyond simply describing statistical techniques. A possible reason may be that measuring IRR/IRA in GT studies is complex due to the iterative nature of the GT process (when and how to apply IRR/IRA during the iterative process of GT) and the multiple coding procedures that GT involves, so researchers have been using ad hoc methods for years, which makes it difficult to use IRR/IRA systematically and extensively. 

This paper presents a process for systematically applying IRR/IRA in GT studies that meets the iterative nature of this qualitative research method and different coding procedures. This helps researchers in developing a shared understanding of the phenomenon being studied by establishing either consistency or agreement among coders, trustworthiness, and robustness of a code frame to co-build a theory through collaborative team science and consortia, and, thus, rigor in qualitative research. To that end, this paper (i) examines the application of IRR/IRA techniques in recent GT studies in software engineering to identify the main challenges and gaps through a systematic literature review, (ii) formalizes a process for systematically applying IRR/IRA in GT, and (iii) shows its feasibility in a GT study. 

The structure of the paper is as follows: Section~\ref{sec:background} provides an overview of the variants of the GT method, the factors that could lead to collaborative coding, criteria for rigorous qualitative research, and the role of IRR/IRA as criteria in qualitative research. Section~\ref{sec:slr} reports a systematic literature review on the application of IRR/IRA techniques in recent GT studies in software engineering. Section~\ref{sec:process} describes a process for systematically applying IRR/IRA in GT and Section~\ref{sec:casestudy} shows feasibility of the process through its application to a GT study. Section~\ref{sec:threats} assesses the validity and reliability of these outcomes. Section~\ref{sec:relatedwork} describes the related work. Finally, conclusions and further work are presented in Section~\ref{sec:conclusion}.

\section{Background}
\label{sec:background}

\subsection{Grounded theory variants}

GT has been defined in its most general form as ``the discovery of theory from data'' (\cite{glaser:1967}, p. 1). GT constitutes a set of different families of research methods, originally rooted in social sciences, but with applications in different domains (psychology, nursing, medicine, education, computer science, managerial and accounting sciences, and even urban planning). Since the publication of its seminal work \cite{glaser:1967}, GT has branched into different families. Most recognizable families within GT are the Classic or Glaserian \cite{glaser:1967}, Straussian \cite{strauss:1990}, and Constructivist or Charmaz GT \cite{charmaz:2014}. They retain a common core of methods, vocabulary, and guidelines, such as coding, constant comparison methods, memo writing, sorting, theoretical sampling, and theoretical saturation, with the final aim of discovering---or developing---a substantive theory grounded in data. They present different nuances in coding procedures that have been referred to as open and initial coding; focused, axial, selective coding; and theoretical coding  \cite{Saldana2012,Kenny:2015}. We can point out at the epistemological underpinnings of GT and the concept of theory sensitivity (namely, the role of literature and academic background knowledge in the process of developing the theory) as the causes of the divergence of schools  \cite{Kenny:2015,STOL:2016}. Hence, the controversial and distinguishing issues of GT can be traced back to the following issues:

\begin{itemize}
    \item Epistemological position: ranging from (naïve) positivism to constructionism. Although the foundational work of GT \cite{glaser:1967} does not adhere to any epistemology, it is acknowledged for its underlying positivist position along with a realist ontology in the classical approach. The Straussian variant modifies this view in favour of a post-positivist position, embracing symbolic interactionism; whereas Charmaz explicitly assumes a constructivist epistemology and a relativist ontology \cite{Kenny:2015}.
    
    \item Theoretical sensitivity is a complex term that in GT denotes both the researcher’s expertise in the research area and his/her ability to discriminate relevant data. The role of literature review is also relevant to grasp this slippery concept of theoretical sensitivity. In Classic GT, the researcher is asked not to be influenced by the existing literature in the construction of the new emerging theory, while being aware of it. Furthermore, the research should approach the data without a clear research question, which should emerge from the data. The Straussian paradigm allows for a much more flexible role of literature review when posing the research question and during the research process, since it will enhance theoretical sensitivity. The Charmazian tradition postulates a much more prominent use of literature to be done at the beginning of the research.     
\end{itemize}

These inconsistencies 
have generated a lot of criticism around GT. Charmaz assumes that the researcher cannot evade from this debate, ``\textit{epistemological stances are, however, significant because they shape how researchers gather their data and whether they acknowledge their influence on these data and the subsequent analysis}'' \cite{Charmaz:2020}; and because it shapes the source of the validity of the obtained knowledge. Thus, the formalization of a new process for GT should be compatible with GT variants abovementioned and flexible enough to different philosophical positions (epistemology and ontology) and theoretical sensitivity.

\subsection{Factors leading collaborative coding}
\label{subsec:factorscollaborativecoding}

GT studies may involve multiple researchers in collaborative coding. According to the Empirical Standards for Software Engineering Research \cite{ralph2021}, in which some authors of this paper were involved, some factors for considering team coding are as follows:

\begin{itemize}
\item Controversiality: the more potentially controversial the judgment, the more multiple raters are needed; e.g.\ recording the publication year of the primary studies in an SLR is less controversial (i.e., coding requires little interpretation) than evaluating the elegance of a technical solution.

\item Practicality: the less practical it is to have multiple rates, the more reasonable a single-rater design becomes; e.g.\ multiple raters applying an a priori deductive coding scheme to some artifacts is more practical than multiple raters inductively coding 2000 pages of interview transcripts.

\item Philosophy: having multiple raters is more important from a realist ontological perspective (characteristic of positivism and falsificationism) than from an idealist ontological perspective (characteristic of interpretivism and constructivism).
\end{itemize}

Because of these factors and the fact that GT studies are becoming larger and more complex, there is a trend toward collaborative coding \cite{erickson:1998,weston:2001,guest:2008}, so formalizing a process for GT studies involving multiple raters can make collaborative science and consortia increasingly systematic and broad.   

\subsection{Criteria for Rigorous Qualitative Research}
\label{subsec:criteriacollaborativecoding}

The appropriate criteria for assessing qualitative research are controversial and often debated not only in the literature \cite{Lincoln:1985,Gibbs:2007,creswell:2017} but also during peer review and dissertation defense. Epistemological and ontological diversity, as well as differences in research traditions between fields, hinders establishing a broad consensus about these criteria. 

Many qualitative researchers claim for \textit{qualitative validity}, which means that the researcher checks for the accuracy of the findings by employing certain procedures, and \textit{qualitative reliability}, which indicates that the researcher’s approach is consistent across different researchers and among different projects \cite{Gibbs:2007,creswell:2017}. In contrast, other qualitative researchers reject validity and reliability altogether in favor of qualitative criteria such as \textit{credibility}, \textit{transferability}, \textit{dependability} (parallel to the conventional criterion of reliability) and \textit{confirmability} \cite{Lincoln:1985}. Therefore, whether to establish reliability in qualitative research and what reliability means for interpretivists building a theory depends on researchers' traditions in different (sub-)disciplines \cite{armstrong:1997,campbell:2013}, from health sciences, psychology, sociology, and business, which may expect formal measures of reliability, to education, information management, and software engineering, which rarely rely on these measures but an increasing interest is emerging in last years \cite{wohlin:2012,nili:2017,mcdonald:2019}. In this regard, it is necessary to analyze the role of IRR/IRA as a criterion in qualitative research.

\subsection{IRR/IRA in qualitative research: general guidelines}

Mcdonald et al. \cite{mcdonald:2019} and O'Connor \& Joffe \cite{oconnor:2020} described norms and guidelines for IRR/IRA in qualitative research in computer and social sciences, respectively. Mcdonald et al. \cite{mcdonald:2019} examined 251 papers in computer science (specifically, in computer-supported cooperative work and human-computer interaction) and found that most papers described a method of IRR or IRA in which two or more raters were involved, and most of the papers used a process that the authors described as inductive. 
O'Connor \& Joffe \cite{oconnor:2020}  conducted an in-depth analysis about arguments in favor of and objections to IRR/IRA in research based on inductive analysis and interpretativist or constructivist epistemology (see Table~\ref{tab:back}). They concluded in words of Braun \& Clarke \cite{braun:2013}, IRR/IRA ``\textit{no necessary imply there is a single true meaning inherent in the data which is the concern underpinning most epistemological objections}'', ``\textit{Rather, it shows that a group of researchers working within a common conceptual framework can reach a consensual interpretation of the data}''.

\begin{table}[h]
    \centering
    \caption{Arguments in favor and against of IRR/IRA in qualitative research. Adapted from \cite{oconnor:2020}}
    \scriptsize
    \begin{tabular}{p{8cm}}
        \hline        Arguments in favor \\\hline
        1. Assess rigor and transparency of the coding frame (refinement)\\
        2. Improve communicability and confidence (beyond an individual interpretation of a researcher)\\
        3. Provide robustness (convergence on the same interpretation of the data)\\ 
        4. Show that analysis is performed conscientiously and consistently \\
        5. Foster reflexivity and dialogue \\
        \hline      Arguments against \\\hline
        1. Contradicts the interpretative epistemological stance\\
        2. Reliability is not an appropriate criterion for judging qualitative work\\
        3. Represents a single, objective, external reality instead of the diversity of interpretations\\
        \hline
    \end{tabular}
    \label{tab:back}
\end{table}
\normalsize

Thus, increasing researchers in different disciplines go beyond IRR/IRA as a statistic or measurement of objectivity and approach IRR/IRA as a tool for improving researcher reflexivity and quality criteria in qualitative research, either for inductive or deductive analysis. 
In fact, Wu et al. \cite{wu:2016} examined various author guidelines for manuscripts reporting qualitative research from a set of journals that include recommendations for IRR/IRA. As reported in Empirical Standards for Software Engineering Research\cite{ralph2021}, in which some authors of this paper were involved, the essential attributes that a study should address when applying IRR/IRA are: 
\begin{itemize}
    \item clearly states what properties were rated
    \item clearly states how many raters rated each property
    \item describes the process by which two or more raters independently rated properties of research objects
    \item describes how disagreements were resolved
    \item indicates the variable type (nominal, ordinal, interval, ratio) of the ratings
    \item reports an appropriate statistical measure of IRR/IRA\footnote{IRR is a correlation measure that can be calculated using Cronbach's $\alpha$, Pearson's $r$, Kendall's $\tau$, and Spearman's $\rho$, among others. IRA is a measure of agreement that can be calculated using Scott's $\pi$, Cohen's $\kappa$, and  and Krippendorff's $\alpha$, among others.}
\end{itemize}

Thus, the formalization of a new process for GT should address the gaps of IRR/IRA in qualitative research.

\section{Systematic Literature Review: A Secondary Study on GT and IRR/IRA}
\label{sec:slr}

This literature review was performed by a team of four researchers (referred to as R1, R2, R3, and R5), who co-authored this article. The objective of this secondary study is to verify whether the the application of IRR/IRA techniques in GT studies is a common practice. Studies have been analyzed in the field of software engineering from 2016 to present. Next, the section is structured following the guidelines by Kitchenham \& Charters \cite{kitchenham2007guidelines} , P\'erez et al. \cite{PEREZ2020110657}, and good practices described in the Empirical Standards for Software Engineering Research \cite{ralph2021}.

\subsection{Planning the SLR}
Planning the SLR consists of developing a review protocol which specifies (i) the review objective and research questions; (ii) the search strategy; (iii) the IC/EC; (iv) the data extraction strategy; and (v) the strategy for synthesizing the extracted data. All these steps are described in the following subsections.

\subsubsection{Review objective \& Research Questions}
This secondary study aims to review the state of the art on the application of IRR/IRA techniques in recent GT studies in software engineering.  The following research questions (RQ) lead this review:

\begin{itemize}
    \item \textbf{RQ1} Have IRR/IRA techniques been applied to GT studies carried out in the field of software engineering?
    \item \textbf{RQ2} How has IRR/IRA been instrumented in previous GT studies in software engineering?
\end{itemize}

\subsubsection{Search Process}\label{ref:search-process}

A formal search strategy is required to find the entire population of scientific papers that may be relevant to answer the research questions. The formal definition of this search strategy allows us to make a replicable and open review of external assessments. The search strategy consists of defining the search space: electronic databases and journals and conference proceedings that are considered key spaces for the review objective. For this work, the search was carried out in the following electronic databases: ScienceDirect, Springer Link, IEEE Xplore, Scopus, and ACM DL. The general search string used for this search is the following
\begin{Verbatim}[breaklines=true]
    ("grounded theory") AND ("inter-rater agreement" OR "inter-rater reliability" OR "inter-judge agreement" OR "inter-judge reliability" OR "inter-coder agreement" OR "inter-coder reliability")
\end{Verbatim}


\begin{table*}[ht]
    \centering
    \caption{Selection criteria}
    \begin{tabular}{p{8cm}|p{8cm}}
    \hline
        \multicolumn{1}{c|}{\textbf{Inclusion criteria}} & \multicolumn{1}{c}{\textbf{Exclusion criteria}} \\\hline
        1. GT should be used as a research methodology in the study & 1. The GT methodology is mentioned but not used in the study\\
        2. IRR/IRA instrument must be calculated and applied in the study & 2. IRR / IRA instrument is mentioned but not used in the study\\
        & 3. IRR/IRA is not applied to the development of the theory during GT coding process (often applied to previous SLR) \\
        & 4. A value for IRR/IRA is provided but the statistical measure is not specified\\\hline
    \end{tabular}
    \label{tab:table2}
\end{table*}

\begin{table*}[ht]
\centering
\caption{Study selection form template}
\begin{tabular}{|l|l|l|l|l|}
\hline
\multicolumn{2}{|l}{Inclusion criteria (of the current iteration)} & & \multicolumn{2}{l|}{Exclusion criteria (of the current iteration)} \\ \hline
\multicolumn{5}{|l|}{Reviewer:}                                                               \\ \hline
Study ID & \multicolumn{2}{l|}{Study title} & Include? (Y/N) & IC/EC \\ \hline
$\ldots$ & \multicolumn{2}{l|}{$\ldots$} & $\ldots$ & $\ldots$ \\ \hline
\end{tabular}
\label{tab:table3}
\end{table*}

\subsubsection{Study Selection Process}
\label{subsec:selection}

The selection process specifies inclusion criteria (IC) and exclusion criteria (EC) to determine whether each potential study should be considered or not for this systematic review (see Table \ref{tab:table2}). Specifically, the study selection process we followed was described in a previous work that aimed to reduce the bias and time spent in the study selection process \cite{PEREZ2020110657}.  

This process was carried out by researchers R2 and R3 (2nd and 3rd authors). They analyzed a set of primary studies to determine whether the study is included or excluded, which is reported in Table \ref{tab:table3}. Both researchers met to contrast their results, refine the IC/EC (if applicable), and calculate the IRA. To calculate IRA, we used Krippendorff’s $\alpha$ (binary) \cite{Krippendorff:2016,Krippendorff:2018} as described in \cite{gonzlezprieto:2020}. When $\alpha \geq 0.8$, the dual selection process is stopped and each researcher independently processes half of the remaining primary studies. However, to ensure that the agreement remains in force (the IC/EC are still interpreted in the same way) and as a quality control measure, some control points were carried out. Using these control points, both researchers reviewed a new set of primary studies and recalculated their agreement.

\subsubsection{Data Extraction Process}

This phase aims to recover the necessary information to answer the research questions. According to the protocol we defined, researchers R1 and R5 (1st and 5th authors) performed data extraction independently and without duplicity (duality is not necessary as the data to be extracted are totally objective). For each primary study we extracted:

\begin{itemize}
    \item Epistemology/Ontology: ranging from positivism or realism/objectivism to  constructivism/interpretivism or relativism.
    \item GT Variant: classic or glausserian, straussian, constructivist/charmazian. If a paper claims to apply a GT approach (mainly coding) but its application is questionable (for instance, the study does not apply theoretical sampling, the constant comparison method, memoing, saturation, or no emerging theories are shown) the study is labelled as ``GT-like approach''.
    \item Data gathering method: semistructured or structured interviews, surveys, etc.
    \item GT coding phases and methods: initial coding, open coding, axial coding, focused coding, selective coding, theoretical coding, constant comparative method, etc.
    \item Did the paper claim to apply Inter-Rater Agreement (IRA) techniques?
    \item Did the paper claim to apply Inter-Rater Reliability (IRR) techniques?
    \item Are Reliability \& Agreement terms correctly used? 
    \item IRA Instrument: Scott's $\pi$, Cohen's $\kappa$, Fleiss's $\kappa$, and Krippendorff's $\alpha$, among others.
    \item IRR Instrument: Pearson's $r$, Kendall's $\tau$, and Spearman's $\rho$, among others.
    \item Process: Brief description of the IRR/IRA process that the authors applied in their GT study.
\end{itemize}

These items are used to define a coding scheme that is processed using a computer-assisted qualitative data analysis (CAQDAS) tool named Atlas.ti v9. 

\begin{figure*}[ht]
\centering
\includegraphics[width=0.9\linewidth]{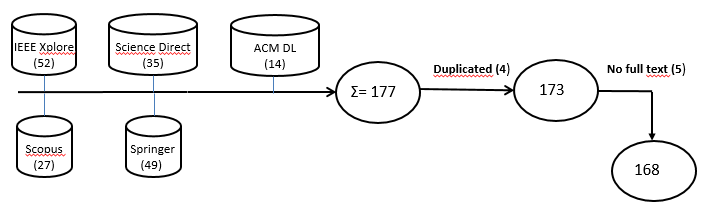}
\caption{Results of the search process}
\label{fig:figure1}
\end{figure*}

\subsubsection{Synthesis Process}

After data extraction, R1 and R3 performed a synthesis process to summarize the main ideas and discoveries from the data. In other words, the synthesis process consists in organizing the key concepts to enable high-order interpretation. 

\subsection{Conducting and Reporting the SLR}

This section reports the results of the study searching, selection, extraction, and synthesis. The systematic review retrieved 173 unduplicated scientific papers. The references of these papers and results of the process are available in a public repository (see Section~\ref{sec:dataavailability}), i.e., a replication package to motivate others to provide similar evidence by replicating this secondary study.

\subsubsection{Results of the Search Process}

Following the review protocol described in Section \ref{ref:search-process}, a search for primary studies was carried out. We located 177 studies from the databases we defined in the protocol, of which 4 were duplicates. Additionally, it was not possible to obtain the full text of 5 studies. Thus, 168 studies were processed (see Figure \ref{fig:figure1}). The search strings used in each electronic database and the 168 studies are listed in the ``replication\_package.xlsx'' file of the repository.

\subsubsection{Results of the Selection Process}
\label{subsec:selectionresults}

\begin{figure}[h]
\centering
\includegraphics[width=9cm]{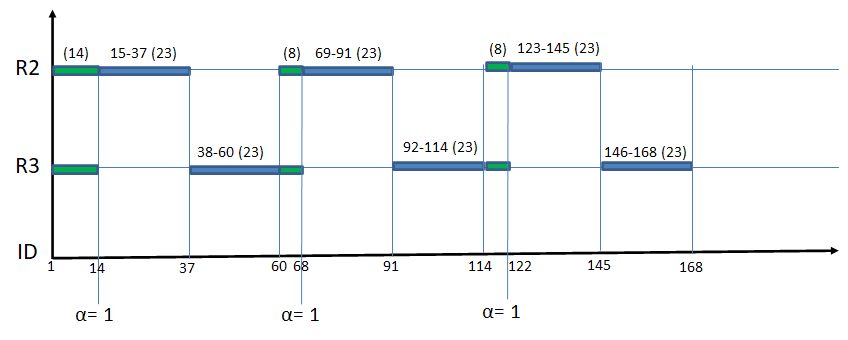}
\caption{Results of inter-rater agreement}
\label{fig:figure2}
\end{figure}

R2 and R3 analyzed 14 (of 168) primary studies based on the IC/EC and obtained the results shown in Table \ref{tab:table5}. These results show that there is no observed disagreement ($D_o = 0$). Therefore, the value of the Krippendorff's coefficient $\alpha = 1 - D_o/D_e = 1$, where $D_e$ is the expected disagreement. This value points out that there exists a very high level of reliability in the selection process. For this reason, the remaining 154 papers are evaluated without dual revision except for the two control points we defined. Hence, to assess ``agreement in force'', R2 and R3 individually reviewed 23 primary studies (i.e., a total of 46). Then, they reviewed 8 primary studies dually and calculated whether the agreement is still in force. This process was repeated (see ``replication\_package.xlsx'' file in the repository). In both control points, a perfect agreement was obtained (see Figure \ref{fig:figure2}). At the end of this process, we selected 49 primary studies, which conducted a GT study and used statistical techniques, either IRR or IRA, for analyzing the consensus when two or more raters were involved in the coding procedures.

\begin{table}[ht]
\centering
\caption{Primary studies included/excluded in the first iteration}
\begin{tabularx}{0.48\textwidth}{
 >{\arraybackslash}X
 >{\arraybackslash}X
 >{\arraybackslash}X
 >{\arraybackslash}X
 >{\arraybackslash}X
 >{\arraybackslash}X
 >{\arraybackslash}X
 >{\arraybackslash}X
 >{\arraybackslash}X
 >{\arraybackslash}X
 >{\arraybackslash}X
 >{\arraybackslash}X
 >{\arraybackslash}X
 >{\arraybackslash}X
 >{\arraybackslash}X
}
\hline
ID & {1} &{2} &{3} &{4} &{5} &{6} &{7} &{8} &{9} &{10} &{11} &{12} &{13} &{14} \\ \hline
R1 & Y & N & N & N & Y & N & Y & Y & N & N & Y & Y & Y & N\\ \hline
R2 & Y & N & N & N & Y & N & Y & Y & N & N & Y & Y & Y & N  \\ \hline
\end{tabularx}
\label{tab:table5}
\end{table}

\subsubsection{Results of the Data Extraction Process}

R1 and R5 performed the data extraction on the 49 primary studies using a coding schema that was implemented using Atlas.ti v9. The results of the process of data extraction are shown in Table \ref{tab:table6}. For the sake of readability, the description of the process whereby the authors of primary studies conducted IRR/IRA in their GT studies is not included, but, for further details, it can be checked in the ``replication\_package.xlsx'' file of the repository.

\subsubsection{Results of the Data Synthesis}
\label{subsec:datasynthesis}

R1 and R3 performed the data synthesis. From the 49 primary studies, we concluded as follows. Most papers did not mention epistemology and ontology except two papers that explicitly mentioned constructivism. Few more than half of the primary studies (28 papers) conducted a GT study as defined in \cite{STOL:2016} and conforms to a set of essential (or desirable) attributes as defined by the Empirical Standards for Software Engineering Research \cite{ralph2021}. From these 28 papers, 16 selected the Straussian variant, 3 papers the Charmaz variant, 1 paper the Classic (or Glaserian) variant, whereas the rest of the papers did not mention any particular GT variant. The other 21 papers conducted a GT-like approach, i.e., the authors only superficially mentioned the GT method to justify the use of coding procedures, without referring to iterative and interleaved rounds of qualitative data collection and analysis for leading to core categories and key patterns, and generating a theory, and without referring to specific GT coding procedures such as, initial, open, focused, axial, selective, and theoretical coding. 

From the 49 primary studies, 33 say to apply IRR whereas 16 IRA. However, according to the formal definitions given by Gisev et al. \cite{gisev2013interrater}, many authors indicated IRR, but instead they applied IRA, i.e, the authors indicated to measure the extent to which raters consistently distinguished between different items on a measurement scale, but instead they measured the extent to which different raters assigned the same precise value for each item being rated. To that end, most of them used Cohen's kappa \cite{cohen1960coefficient} and Krippendorff's $\alpha$ \cite{Krippendorff:2018}. Thus, it seems clear that IRR and IRA are frequently interchangeable terms. 

\onecolumn
\begin{table*}
\tiny
\begin{tabular}{|l|l|l|l|l|c|c|c|l|}
\hline \textbf{ID} & \multicolumn{1}{c|}{\textbf{\begin{tabular}{c}Epistemology\\ Ontology\end{tabular}}} & \multicolumn{1}{c|}{\textbf{GT variant}} & \multicolumn{1}{c|}{\textbf{Recollection  method}} & \multicolumn{1}{c|}{\textbf{GT Coding Phases}} & \textbf{IRA} & \textbf{IRR} & \textbf{\begin{tabular}{c}Correctly \\	used?\end{tabular}} & \multicolumn{1}{c|}{\textbf{Coefficient}} \\   \hline
\hline
01   & Not mentioned   & Straussian GT      & Interviews data     & \begin{tabular}[c]{@{}l@{}}Open coding\\ Axial coding\end{tabular}    & -- & IRR  & No (IRA)   & Cohen’s $\kappa$     \\ \hline
05   & Not mentioned   & Straussian GT      & Video data   & Not mentioned      & -- & IRR  & No (IRA)   & Fleiss $\kappa$    \\ \hline
07   & Not mentioned   & Straussian GT      & Literature      & \begin{tabular}[c]{@{}l@{}}Open coding\\ Axial coding\\ Selective coding\\ Const.\ comparison\end{tabular} & -- & IRR  & No (IRA)   & Krippendorff's $\alpha$      \\ \hline
08   & Not mentioned   & \begin{tabular}[c]{@{}l@{}}Straussian GT\\ (inductive analysis)\end{tabular} & \begin{tabular}[c]{@{}l@{}}Screen recordings \\ Survey data \\ Interviews data\end{tabular}  & \begin{tabular}[c]{@{}l@{}}Open coding\\ Affinity diagramming \\ Memoing\end{tabular}      & -- & IRR  & No (IRA)   & Fleiss’s $\kappa$    \\ \hline
11   & Not mentioned   & Inductive analysis   & Twitter data      & Open coding        & -- & IRR  & No (IRA)   & $\kappa$ (unknown version)  \\ \hline
12   & Not mentioned   & Straussian GT      & Interviews data     & \begin{tabular}[c]{@{}l@{}}Open coding\\ Axial coding\end{tabular}    & -- & IRR  & No (IRA)   & Cohen’s $\kappa$  \\ \hline
13   & Not mentioned   & Not mentioned      & Interviews data     & Axial coding       & IRA  & -- & Yes    & Cohen’s $\kappa$  \\ \hline
17   & Not mentioned   & Not mentioned      & \begin{tabular}[c]{@{}l@{}}Video data\\ Text specifications\end{tabular}    & Not mentioned      & -- & IRR  & No (IRA)   & \begin{tabular}[c]{@{}l@{}}Percent agreement\\ Cohen’s $\kappa$\end{tabular}     \\ \hline
24   & Not mentioned   & Classic GT    & Discourse files     & \begin{tabular}[c]{@{}l@{}}Open coding\\ Axial coding\\ Selective coding\end{tabular} & -- & IRR  & No (IRA)   & $\kappa$ (unknown version) \\ \hline
26   & Constructivism     & Straussian GT      & Survey data  &    & -- & IRR  & No (IRA)   & Krippendorff’s $\alpha$ \\ \hline
28   & Not mentioned   & Straussian GT      & Survey data  & Open coding        & -- & IRR  & No (IRA)   & Krippendorff’s $\alpha$   \\ \hline
31   & Not mentioned   & Not mentioned      & Focus group results      & Not mentioned      & -- & IRR  & Yes  & Not mentioned     \\ \hline
34   & Not mentioned   & Straussian GT      & Interviews data     & Open coding        & -- & IRR  & No (IRA)   & Cohen’s $\kappa$     \\ \hline
37   & Not mentioned   & Straussian GT      & Interviews data     & Open coding        & IRA  & -- & Yes  & Cohen’s $\kappa$     \\ \hline
41   & Not mentioned   & GT-like approach     & Logs     & \begin{tabular}[c]{@{}l@{}}Open coding\\ Thematic analysis\end{tabular}    & IRA  & -- & Yes  & Cohen’s $\kappa$     \\ \hline
44   & Not mentioned   & GT-like approach     & Interview data & Open coding        & IRA  & -- & Yes  & Cohen’s $\kappa$     \\ \hline
55   & Not mentioned   & GT-like approach     & Literature      & Not mentioned      & -- & IRR  & Yes  & Kendall’s    \\ \hline
57   & Not mentioned   & GT-like approach     & Interview data      & Inductive analysis      & -- & IRR  & No (IRA)   & Cohen’s $\kappa$     \\ \hline
58   & Constructivism     & Charmaz GT    & \begin{tabular}[c]{@{}l@{}}Interview data\\ Survey data\end{tabular}   & \begin{tabular}[c]{@{}l@{}}Inductive analysis\\ Memoing\end{tabular}     & IRA  & -- &  Yes & Cohen’s $\kappa$     \\ \hline
59   & Not mentioned   & GT-like approach     & Literature data     &    & IRA  & -- &  Yes & Krippendorff’s $\alpha$   \\ \hline
60   & Not mentioned   & Straussian GT      & \begin{tabular}[c]{@{}l@{}}Text specifications \\ (users reviews) \end{tabular}    & \begin{tabular}[c]{@{}l@{}}Open coding\\ Axial coding\\ Selective coding\end{tabular} & IRA  & -- & Yes  & Cohen’s $\kappa$     \\ \hline
61   & Not mentioned   & GT-like approach     & Text specifications      &    & IRA  & -- &   & Fleiss’s $\kappa$    \\ \hline
62   & Constructivism     & Straussian GT      & Survey data  & Not mentioned      & IRA  & -- & Yes  & Cohen’s $\kappa$     \\ \hline
65   & Not mentioned   & Charmaz GT    & \begin{tabular}[c]{@{}l@{}} Text specifications \\ (functional requirements) \end{tabular} & Not mentioned      & IRA  & -- & Yes  & Cohen’s $\kappa$     \\ \hline
66   & Not mentioned   & Straussian GT      & \begin{tabular}[c]{@{}l@{}}Video data\\  Text data (comments)\end{tabular}   & Not mentioned      & IRA  & -- &  Yes & Cohen’s $\kappa$  \\ \hline
67   & Not mentioned   & Straussian GT      & Interview data      & \begin{tabular}[c]{@{}l@{}}Open coding\\ Axial coding\\ Selective coding\\ Const.\ comparison\end{tabular} & -- & IRR  & No (IRA)   & Cohen’s $\kappa$     \\ \hline
74   & Not mentioned   & GT-like approach     & Interview data      & Not mentioned      & -- & IRR  & No (IRA)   & Cohen’s $\kappa$     \\ \hline
78   & Not mentioned   & GT-like approach     & Interview data      & Content Analysis   & -- & IRR  & No (IRA)   & Cohen’s $\kappa$     \\ \hline
83   & Not mentioned   & GT-like approach     & Survey data  & Not mentioned      & -- & IRR  & No (IRA)   & Cohen’s $\kappa$     \\ \hline
87   & Not mentioned   & Charmaz GT    & \begin{tabular}[c]{@{}l@{}}Case studies data: \\ Text specifications \\ and images (diagrams)\end{tabular}  & Memoing        & IRA  & -- &  Yes & Cohen’s $\kappa$     \\ \hline
88   & Not mentioned   & Straussian GT      & Survey data  & \begin{tabular}[c]{@{}l@{}}Open coding\\ Axial coding\end{tabular}    & -- & IRR  & No (IRA)   & Krippendorff’s $\alpha$    \\ \hline
89   & Not mentioned   & GT-like approach     & Interview data      & \begin{tabular}[c]{@{}l@{}}Content analysis\\ Thematic analysis\end{tabular}      & -- & IRR  & No (IRA)   & Cohen’s $\kappa$     \\ \hline
92   & Not mentioned   & \begin{tabular}[l]{@{}l@{}}GT but variant\\ is not specified\end{tabular}  & Instagram data      & \begin{tabular}[c]{@{}l@{}}Open coding\\ Axial coding\\ Const.\ comparison\end{tabular}     & IRA  & -- &  Yes & Cohen’s $\kappa$     \\ \hline
97   & Not mentioned   & \begin{tabular}[l]{@{}l@{}}GT but variant\\ is not specified\end{tabular}  & Interview data      & \begin{tabular}[c]{@{}l@{}}Initial coding\\ Open coding\\ Axial Coding\end{tabular}   & -- & IRR  & No (IRA)   & \begin{tabular}[c]{@{}l@{}}Cohen’s $\kappa$\\ Scott’s $\pi$ \end{tabular} \\ \hline
102  & Not mentioned   & GT-like approach     & \begin{tabular}[c]{@{}l@{}}Logs\\ Text specifications\end{tabular} & \begin{tabular}[c]{@{}l@{}}Initial coding\\ Axial coding\end{tabular}      & -- & IRR  & No (IRA)   & Cohen’s $\kappa$     \\ \hline
110  & Not mentioned   & GT-like approach     & Audio data   & Not mentioned      & -- & IRR  & No (IRA \& IRR) & \begin{tabular}[c]{@{}l@{}}Cohen’s $\kappa$\\ Shaffer’s $\rho$\end{tabular}     \\ \hline
111  & Not mentioned   & 
     GT-like approach   & \begin{tabular}[l]{@{}l@{}}Interview data\\ Focused group data\end{tabular} & Not mentioned      & -- & IRR  & No (IRA)   & Cohen’s $\kappa$     \\ \hline
128  & Not mentioned   & GT-like approach  & Image data   & Open coding        & -- & IRR  & No (IRA)   & Cohen’s $\kappa$     \\ \hline
132  & Not mentioned   & GT-like approach     & Text specifications      & Not mentioned      &   & IRR  & No (IRA)   & \begin{tabular}[c]{@{}l@{}}Percent agreement\\ Fleiss’ $\kappa$\end{tabular} \\ \hline
138  & Not mentioned   & Straussian-GT    & Text specification  & \begin{tabular}[c]{@{}l@{}}Open coding\\ Axial coding\\ Selective coding\\ Content analysis\end{tabular}     & IRA  & -- & Yes  & Cohen’s $\kappa$     \\ \hline
140  & Not mentioned   & \begin{tabular}[l]{@{}l@{}}GT but variant\\ is not specified\end{tabular} & \begin{tabular}[c]{@{}l@{}}Interview data\\ Text specifications\end{tabular}  & Content analysis   & -- & IRR  & No (IRA)   & Kripendorff’s $\alpha$    \\ \hline
144  & Not mentioned   & GT-like approach     & Interview data      & Not specified      & -- & IRR  & No (IRA)   & Cohen’s $\kappa$     \\ \hline
147  & Not mentioned   & GT-like approach     & Literature data     & Not specified      & -- & IRR  & No (IRA)   & Cohen’s $\kappa$     \\ \hline
151  & Not mentioned   & GT-like approach     & Text specifications      & Not specified      & -- & IRR  &   & Pearson‘s $r$   \\ \hline
153  & Not mentioned   & Straussian-GT    & Text specifications      & \begin{tabular}[c]{@{}l@{}}Open coding\\ Axial coding\\ Selective coding\end{tabular} & -- & IRR  & No (IRA)   & Cohen’s $\kappa$     \\ \hline
154  & Not mentioned   & GT-like approach     & Text specifications      & Not specified      & IRA  & -- &   & Fleiss’ $\kappa$     \\ \hline
158  & Not mentioned   & \begin{tabular}[l]{@{}l@{}}GT but variant\\ is not specified\end{tabular}  & Text specifications      & \begin{tabular}[c]{@{}l@{}}Open coding\\ Axial coding\\ Selective coding\\ Content analysis\end{tabular}     & -- & IRR  & No (IRA)   & $\kappa$ (unknown version)  \\ \hline
161  & Not mentioned   & GT-like approach     & Interview data      & Not mentioned      & IRA  & -- &   & Percent agreement      \\ \hline
163  & Not mentioned   & GT-like approach     & \begin{tabular}[c]{@{}l@{}}Interviews data\\ Surveys (questionnaires)\end{tabular} & \begin{tabular}[c]{@{}l@{}}Open coding\\ Axial coding\end{tabular}    & -- & IRR  & No (IRA)   & Cohen’s $\kappa$     \\ \hline
\end{tabular}

\caption{Data extraction}
\label{tab:table6}
\end{table*}

\twocolumn
\normalsize

In addition to this descriptive analysis, an analytical reasoning of the IRA/IRR application in these GT studies shows some lacks:

\begin{itemize}

    \item None of the primary studies provide a reasonable justification for collaborative coding in terms of controversiality, practicality, or philosophy (see Section~\ref{subsec:factorscollaborativecoding}). 
    
    \item However, all of them justify collaborative coding and the use of IRR/IRA as criteria for rigorous qualitative research (see Section~\ref{subsec:criteriacollaborativecoding}), i.e., as a means for avoiding researcher bias (e.g., ID7, ID55); gaining in reliability and consensus (e.g., ID67, ID110, ID151) and robustness (e.g., ID17); testing validity (e.g., ID67, ID110, ID132, ID144); and for refining codebooks (e.g., ID8, ID34) and clarifying definitions (e.g., ID12).
    
    \item Almost no primary studies describe the process by which two or more raters independently rated codes in sufficient  detail, for example, who defines the code frame, what the code frame is, the units of coding (i.e. how data is segmented into meaningful quotes, e.g.\, a paragraph, a line, etc.), 
    how disagreements were resolved, etc. Only two primary studies (ID8 and ID55) mentioned an iterative process, which is described in little more than a paragraph. 
    
    \item Only 8 primary studies explicitly indicated the corpus (data) over which IRR/IRA is applied.
    
    \item Only 5 primary studies explicitly indicate the minimum threshold that indicates acceptable reliability/agreement. 
    
    \item Most papers report an inappropriate statistical measure of IRR/IRA; hence, 33 primary studies stated to use Cohen's kappa and Krippendorff's $\alpha$ to measure IRR when these statistical techniques measure IRA. 
    
\end{itemize}

Next, we show some excerpts of the primary studies to make explicit the chain of evidence. 
Hence, ID144 and ID161 pointed out at the use of quantitative techniques to \textit{enhance the quality} of qualitative data:  

\vspace{0.2cm}
\noindent \fbox{\begin{minipage}{8.5cm}
\textit{ID144 "To enhance the rigor of the quantitative analysis of qualitative data analysis, a triangulation of analysts [16] was employed. Two researchers coded four interviews with the nine care transition outcomes separately; they met and reviewed their coding to discuss differences and refine outcome definitions. The two researchers then coded two more interviews separately to evaluate inter-rater reliability to strengthen the internal validity of the research."}
\end{minipage}}
\vspace{0.2cm}

ID01, ID11, ID26, ID37, ID58, ID62, ID66, ID74, ID83, ID88, and ID163 superficially described the IRR/IRA process by indicating only the number of coders, and vaguely how disagreements were discussed.  

\vspace{0.2cm}
\noindent \fbox{\begin{minipage}{8.5cm}
\textit{ID01 “Coding was performed independently by two coders who met frequently to discuss codes in order to ensure high inter-rater reliability.”}
\end{minipage}}
\vspace{0.2cm}

\vspace{0.2cm}
\noindent \fbox{\begin{minipage}{8.5cm}
\textit{ID11 “Each tweet was coded independently by two coders. Kappa coefficients measuring inter-coder reliability above chance agreement ranged from fair to good (50\% to 88\%).”}
\end{minipage}}
\vspace{0.2cm}

ID144 and ID153 acknowledged a minimum threshold for achieving acceptable agreement when using Cohen's $\kappa$ and ID88 described an acceptance threshold for Krippendorff’s alpha at $>$ 0.8. However, only ID07 and ID57 seem to describe a minimum threshold of Krippendorff’s $\alpha$ as a tool for progressively improving the consistency of a code frame, although it is not explicitly described. 

\vspace{0.2cm}
\noindent \fbox{\begin{minipage}{8.5cm}
\textit{ID144 "Cohen’s kappa was calculated for all outcomes; all values were above the acceptable value of 0.8 and indicate that the interpretation and coding of interview data are reliable.”}
\end{minipage}}
\vspace{0.2cm}

\vspace{0.2cm}
\noindent \fbox{\begin{minipage}{8.5cm}
\textit{ID07 “The results of the Krippendorff’s  $\alpha$ test suggest that there was a 69\% agreement between the observers. Because this result was below the commonly accepted threshold of an $\alpha$ of 80\%, the first two authors deliberated over the differences to form one consistent initial coding set.”}
\end{minipage}}
\vspace{0.2cm}

\vspace{0.2cm}
\noindent \fbox{\begin{minipage}{8.5cm}
\textit{ID153 “A further independent inter-rater test was performed which achieved a 75\% agreement which according to Landis and Koch is a “substantial agreement”.”}
\end{minipage}}
\vspace{0.2cm}

ID05, ID13, ID17, ID24, ID28, ID89, ID92, and ID128 described the corpus over which IRA is calculated (25\%, 20\%, approx. 30\%, 10\%, approx. 10\%, 26\%, 10\% and 20\% respectively). Specifically, ID17, although uses IRR to refer to IRA, indicates the role of the researchers and the percentage of data over which IRA is calculated.

\vspace{0.2cm}
\noindent \fbox{\begin{minipage}{8.5cm}
\textit{
ID17 “The bulk of the coding was performed by the first author. In order to ensure the robustness of the coding system, the remaining three authors performed two independent coding passes of a subset of 50 of the 230 artifacts in the first pass, and 25 of the 230 artifacts and 6 of the 60 videos, at two stages in the development of the code books. We calculated the inter-coder reliability ratio as the  number of agreements divided by the total number of codes [...]”} 
\end{minipage}}
\vspace{0.2cm}

\vspace{0.2cm}
\noindent \fbox{\begin{minipage}{8.5cm}
\textit{
ID128 To test the inter-coder reliability, the primary researcher coded all 154 records, and subsequently the second coder coded every fifth record in the dataset. Cohen’s Kappa coefficient was found to be 0.84, indicating high agreement between the two coders}
\end{minipage}}
\vspace{0.2cm}

ID55, ID59, ID65, and ID89 are the only ones of the few to lightly describe the expertise of raters involved in the coding process, but none of them mention a specific training on the coding process. 

\vspace{0.2cm}
\noindent \fbox{\begin{minipage}{8.5cm}
\textit{ID55 “To avoid the researcher’s bias, we have performed an inter-rater reliability test between mapping team and indented experts [...]”}
\end{minipage}}
\vspace{0.2cm}

\vspace{0.2cm}
\noindent \fbox{\begin{minipage}{8.5cm}
\textit{ID59 “The initial team of paper taggers was made up of seven post-docs and graduate students with some association to the University of Trento and some experience with goal modeling.”}
\end{minipage}}
\vspace{0.2cm}

\vspace{0.2cm}
\noindent \fbox{\begin{minipage}{8.5cm}
\textit{ID02 “A total of 433 excerpts were extracted from the interview transcripts (excluding answers from Q4 which was quantitative), and 113 excerpts (26\%) were randomly selected and double coded by two independent coders who were social science graduate students.”}
\end{minipage}}
\vspace{0.2cm}

ID110 relies on automated text coding. Data is codified by an algorithm and subsequently validated by  human experts. 

\vspace{0.2cm}
\noindent \fbox{\begin{minipage}{8.5cm}
\textit{ID110 “To code the data, we developed an automated coding scheme using the nCodeR package for the statistical programming language R [...]. We used nCodeR to develop automated classifiers for each of the codes in Table 1 using regular expression lists [...]. To create valid and reliable codes, we assessed concept validity by requiring that two human raters achieve acceptable measures of kappa and rho, and reliability by requiring that both human raters independently achieve acceptable measures of kappa and rho compared to the automated classifier.”}
\end{minipage}}
\vspace{0.2cm}

ID 67 is the one on describing the IRR/IRA process in relation to the different GT coding procedures (i.e., open, axial, and selective coding phases). However, most of the papers use IRR/IRA as a finalist measure like ID87. Hence, ID97 and ID153 explicitly indicate that IRA is calculated after the theoretical saturation was reached. 

\vspace{0.2cm}
\noindent \fbox{\begin{minipage}{8.5cm}
\textit{ID67 “Hence, in order to implement IRR, two coders were involved in independent analysis and coding the transcripts from the interviews and the convergence of their findings was evaluated at the end of each open, axial, and selective coding phases. In cases of conflicts between the decisions made by these two coders, a third coder was involved in the discussions for resolving the conflicts. At the end of each coding phase, we merged the coding files from ATLAS.ti and exported the coding results of each researcher into Microsoft Excel. We used Microsoft Excel to calculate Kappa as a measure of IRR..”}
\end{minipage}}
\vspace{0.2cm}

\vspace{0.2cm}
\noindent \fbox{\begin{minipage}{8.5cm}
\textit{ID87 “After developing the coding scheme through grounded theory as described above, we conducted a second phase of analysis to test inter-rater reliability..”}
\end{minipage}}
\vspace{0.2cm}

\vspace{0.2cm}
\noindent \fbox{\begin{minipage}{8.5cm}
\textit{ID153 "Selective coding is the final coding process in GTM, and involves the selection of core categories of the data. Selective coding systematically relates the categories identified in axial coding, and integrates and refines them to derive theoretical concepts.
After theoretical saturation, we conducted an inter-rater reliability test evaluation using Cohen’s kappa.”}
\end{minipage}}
\vspace{0.2cm}

Only ID08 and ID34 explicitly mentioned an iterative process that aims to improve a code frame (e.g.\ removing ambiguous codes) and, thus, improve researchers' reflexivity.

\vspace{0.2cm}
\noindent \fbox{\begin{minipage}{8.5cm}
\textit{ID08 “Four coders independently coded two samples to refine the coding scheme. We then discussed and used affinity diagramming to synthesize emerging themes. Next, we went through several iterations to check another two samples individually. The purpose of this step was to confirm the legitimacy of the coding scheme and to check the inter-rater reliability. After several iterations, four coders reached a suitable level of agreement (Fleiss’s kappa, $\kappa = 0.71$)..”}
\end{minipage}}
\vspace{0.2cm}

\vspace{0.2cm}
\noindent \fbox{\begin{minipage}{8.5cm}
\textit{ID34 “Initially, the 1st and 2nd author each independently coded a new sample of five analyses (20\% of the data), receiving a low Cohen’s Kappa of 0.55 [37]. Both authors discussed disagreements, refined the code book, and repeated the process on a new sample of five interviews. With a moderate Cohen’s Kappa of 0.71 [37], the two authors labeled all remaining interviews together, allowing for multiple labels where needed, as decided through discussion and consensus. Afterwards, our analysis followed ‘data-driven’ thematic analysis [42] where we clustered our coded data into themes.”}
\end{minipage}}
\vspace{0.2cm}

\section{A Process for IRR/IRA in Grounded Theory Studies}
\label{sec:process}

This section presents a process for systematically applying IRR/IRA in GT studies in which multiple researchers are involved in collaborative coding. Before describing this process, it is necessary to highlight two concerns that the process should consider. The first one is about coding. As McDonald et al. \cite{mcdonald:2019} examined in previous literature, coding is sometimes used to describe a process of \textit{inductive interpretation}, and other times is used to describe a process of \textit{deductive labeling} of data with preexisting codes, even sometimes both approaches are integrated as Cruzes and Dyba recommended for thematic analysis in software engineering \cite{cruzes:2011}.

The second one is about the purpose of measuring IRR/IRA. When multiple raters collaboratively code, consensus could be reached through ``intensive group discussion, dialogical intersubjectivity, coder adjudication, and simple group consensus as an agreement goal'' \cite{Saldana2012}. However, \textit{you cannot improve what you cannot measure}, and precisely, IRR and IRA techniques allow researchers to measure consistency and agreement across multiple coders. Measuring consistency and agreement among raters, where appropriate, promotes ``systematicity, communicability, and transparency of the coding process; reflexivity and dialogue within research teams; and helping to satisfy diverse audiences of the trustworthiness of the research'' \cite{oconnor:2020}. It is particularly crucial to identify mistakes before the codes are used in developing and testing a theory or model, i.e., to ensure robustness before analyzing and aggregating the coding data. 
Weak confidence in the data only leads to uncertainty in the subsequent analysis and generates doubts on findings and conclusions. In Krippendorff's own words: ``If the results of reliability testing are compelling, researchers may proceed with the analysis of their data. If not, doubts prevail as to what these data mean, and their analysis is hard to justify'' \cite{Krippendorff:2018}. 

Thus, IRR and IRA provide a key tool for achieving inter-coder consistency and agreement by encouraging consensus and reflexivity \cite{hammer:2014} and a shared understanding of the data, discovering where coders disagree, and revealing weaknesses in coding definitions \cite{mcdonald:2019}, overlaps in meaning \cite{MacPhail:2016}, or difficulties in consensus given the nature of data \cite{hammer:2014}. For us, the process of reaching consensus, either consistency or agreement (or both), is more important than its measurement, although measurement is the key to conducting this process. 

\subsection{The Process}
The process for GT studies we propose (see Figure~\ref{fig:figure3}) starts with an initial research question(s) and data collection, either via purposive, convenience, or theoretical sampling strategies. In this process, N$>$1 raters are involved, which depends on the size of the GT study or its controversiality. What is clear is that the greater the number of raters (coders), the more difficult it is to reach a consensus (either consistency IRR or agreement IRA), but the trustworthiness is improved. Next, the raters are involved in the coding of a subset of data (e.g.\ interviews, qualitative survey responses, or any other data subject to qualitative analysis) as follows: 

\begin{figure*}[h!]
\centering
\includegraphics[width=11cm]{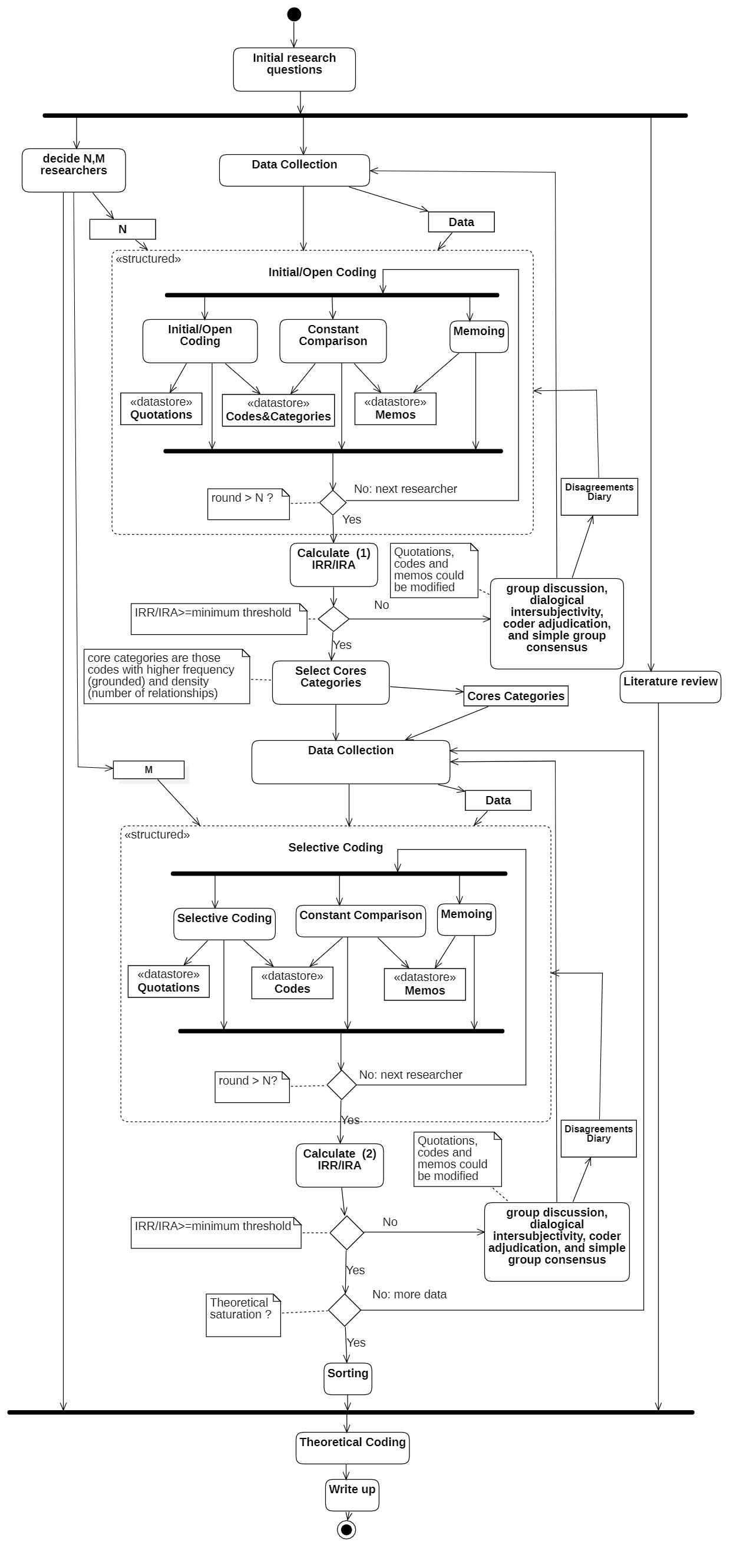}
\caption{Process for using IRR/IRA in GT studies (UML Activity diagram)}
\label{fig:figure3}
\end{figure*}

\textbf{Initial/open coding}: This activity involves multiple rounds of coding, constant comparison, and memoing (see Figure~\ref{fig:figure3}), specifically one round per rater involved in the collaborative coding. The first rater analyzes the subset of selected data (e.g., between 5-10 instances\footnote{This is an arbitrary number selected by the researcher, which depends on the quantity and quality of the data collected and the availability of human resources.}) by reviewing the data line by line, creating quotations (highlighted segments of text), assigning new codes to the quotations, and writing memos, i.e., notes about ideas or concepts potentially relevant to the research. As more data instances are analyzed, the resulting codes (code frame or codebook) are refined by using the constant comparison method that forces the rater to go back and forth. The following raters analyze the same subset of selected data in which the quotations that the previous raters created are visible (however, raters never see the coding of previous raters, i.e., the codes assigned to each quotation), creating new quotations if necessary (for relevant data that were omitted by previous raters), assigning new codes or previously defined codes to the quotations, and writing new memos if necessary, while constantly comparing codes with each other, within the data instance and between instances. This process is, thus, an integrated approach of inductive and deductive coding. 

Once all raters have coded the subset of selected data, the collaborative coding is the input for measuring IRR and/or IRA (see activity \textbf{Calculate (1) IRR/IRA} in Figure~\ref{fig:figure3}). 

\begin{itemize}
    \item  If this measure is less than a minimum threshold (that could vary depending on the statistical technique), a group discussion to collaboratively reach consensus and/or agreement is followed (see activity \textbf{group discussion, dialogical inter-subjectivity, coder adjudication, and simple group consensus} in Figure~\ref{fig:figure3}). This activity aims to identify coding disagreements, weaknesses in coding definitions, overlaps in meaning, etc. as mentioned before, which are documented in a \textbf{\textit{disagreements diary}}. After possible modifications on quotations, codes, and memos, a new iteration of initial/open coding starts over a new subset of selected data (if necessary, new data are collected).
    
    \item If this measure is higher or equal to the minimum threshold, the following activity starts.  
\end{itemize}

\textbf{Selection of core categories} aka. variables (see Figure~\ref{fig:figure3}): raters select core categories from the most relevant and important codes obtained in the previous coding procedure (the usual criteria to select core categories can be frequency and density, i.e., the number of code occurrences and the number of relationships among codes).

\textbf{Selective coding}: This activity also involves multiple rounds of coding, constant comparison, and memoing (see Figure~\ref{fig:figure3}), specifically one round per rater involved in the collaborative coding. All raters analyze the same new subset of selected data by reviewing the data line by line, creating quotations (segments of text), assigning them a core category (i.e, subcodes of a core category), writing memos, and comparing the categories with one another. This coding procedure is an integrated approach of inductive and deductive coding that only focuses on the core categories and subcodes of these categories. Thus, coding is a deductive process of labeling data with preselected core categories and an inductive process of creating and labeling data with possible new subcodes of these preselected core categories. Again, successive raters analyze the same subset of selected data in which the quotations that the previous raters created are visible (however, raters never see the coding of previous raters, i.e., the codes assigned to each quotation).

Once all raters have coded the subset of selected data, the collaborative coding is the input for measuring IRR and/or IRA (see activity \textbf{Calculate (2) IRR/IRA} in Figure~\ref{fig:figure3}).  

\begin{itemize}
    \item  If this measure is less than a minimum threshold (that could vary depending on the statistical technique), a group discussion to collaboratively reach consensus and/or agreement is followed  (see activity \textbf{group discussion, dialogical inter-subjectivity, coder adjudication, and simple group consensus} in Figure~\ref{fig:figure3}). This process aims to identify coding disagreements, weaknesses in categories definitions, overlaps in meaning, etc. as mentioned before, which are documented in a \textbf{\textit{disagreements diary}}. After possible modifications on quotations, core categories and memos, a new iteration of selective coding starts over another subset of selected data (if necessary, new data are collected).
    
    \item If this measure is higher or equal to the minimum threshold, researchers evaluate if theoretical saturation is reached. If not, new data are collected via theoretical sampling and a new iteration of selective coding starts. 
    
    \item If theoretical saturation is reached, researchers are involved in the procedures of sorting core categories and memos and theoretical coding, until building and writing up a theory (see Figure~\ref{fig:figure3}).
\end{itemize}

Finally, the role of the literature has been formalized to be compatible with the three GT variants, either reviewing the literature to fit the purpose of the GT study \cite{charmaz:2014} or delaying literature review until the theory has emerged to validate it \cite{glaser:1967}. 

The process here described meets the iterative nature of the GT research method and helps in developing a shared understanding of the phenomenon being studied by establishing either consistency or agreement among coders, and thus, the trustworthiness and robustness of a code frame to co-build a theory through collaborative team science and consortia.

\subsection{Discussion}

Next, some aspects related to the process are described and discussed:

1. The process does not state anything about how to code. It does state the phases, activities, and milestones in a GT process that incorporates IRR/IRA. Figure~\ref{fig:figure3} indicates where and when to use these statistics, but says nothing about how to do open coding, axial, or theoretical coding.

2. The process does not impose any restrictions about how many researchers collect data or who collect these data, i.e., there can be multiple data collectors or only one, and they can be the coders themselves or different researchers. This flexibility is one of the benefits of the proposed process. Moreover, the number and identity of the coders are allowed to be dynamically changed if the statistic used to measure the IRR/IRA allows it (e.g., Krippendorff coefficients do). Hence, the N and M values (Figure~\ref{fig:figure3}) can be the same or different and refer to the same or different coders.

3. According to the process here described, the coders act sequentially, i.e., one coder does not start his/her coding process until the previous one has finished and he/she uses the quotations and the code frame (or codebook) generated so far (although the coding of previous coders is not visible). However, perhaps it is worth considering that parallel coding would save time. Thus, why not code in parallel? If several coders work simultaneously, each coder could define a different set of codes from a morphological, lexical, syntactic, or semantic point of view.  Only the latter case is relevant when building a theory since the disagreement would be about the constructs themselves (their meaning). The other sources of disagreement only imply loss of time in meetings to work out the disagreements. Hence, a morphological (ball versus balls) or lexical (kids versus children) disagreement does not involve a disagreement on the meaning of the construct but only on the way of referring to it. The same applies to syntactic disagreements; that is, the meaning of a code can be expressed with a phrase that admits a different order of its constituent parts or with semantically equivalent phrases. To avoid having to resolve these kinds of ``format disagreements'', we propose that coders work sequentially by using the codes and quotations defined by previous coders. This process does not prevent the generation of "format disagreements", but it does avoid an initial explosion of codes (and disagreements) that must be agreed.

4. GT methodology prescribes an iterative process that ends when saturation is reached. When a single researcher is involved in the process, this condition is necessary and sufficient. However, when there are several coders, saturation is necessary but not sufficient. What happens if saturation is reached and there is no agreement among the coders? If we go on to build the theory with no agreement on the semantics of the constructs, we may lose the benefits we are pursuing with the use of IRR/IRA. The opposite case---agreement is reached but not saturation---implies new iterations until both are reached.

\section{Application of the Process for IRR/IRA in a GT Study}
\label{sec:casestudy}

This section describes part of a GT study that the authors have been conducting in the domain of \textit{Edge Computing} and \textit{DevOps} in industry (\textit{EdgeOps}) over the last while, which aims to illustrate the application of the proposed process for IRR/IRA in GT studies. This process has involved simultaneous data collection and analysis as described in Section~\ref{sec:process}. 

According to the \textit{purposive sampling strategy}, we initially collected data from a set of participants from leader organizations in the Internet of Things domain, which are currently committee members of the Master's Degree in Distributed and Embedded Systems Software\footnote{\url{http://msde.etsisi.upm.es/}} and Master's Degree in IoT\footnote{\url{https://masteriot.etsist.upm.es/?lang=en}} at Universidad Polit\'ecnica de Madrid, Spain. Then, we moved on to \textit{theoretical sampling} and iteratively collected more data based on the concepts or categories that were relevant for the emerging theory until ICA value exceeded a given threshold and theoretical saturation was reached. A total of 27 answers were collected from a open-ended questionnaire available in \url{https://es.surveymonkey.com/r/PMWD7ZM}. 

This GT study involved three researchers in the coding process (denoted by R1, R2, and R3) because of the controversiality of the terms around EdgeOps, whose definition, characterization, benefits, implications, and challenges have little consensus among the community due to its novelty. As multiple coders were involved, we applied inter-coder agreement (ICA), and specifically Krippendorff coefficients \cite{Krippendorff:2018,gonzlezprieto:2020}, to improve the quality of our qualitative analysis---i.e., discover disagreements and reveal weaknesses in coding definitions, overlaps in meaning, etc.---and gain in researchers' reflexivity and a shared understanding of the data. The qualitative analysis was also supported by the tool Atlas.ti v9, which includes specific functionality to calculate Krippendorff coefficients.

Next subsections describe the main notions about Krippendorff coefficients we have used, and the multiple iterations that have been necessary during both initial/open coding and selective coding to exceed a certain threshold---that the community has approved---and during which, codes and categories (and memos) were improved and clarified as disagreements revealed weaknesses in coding definitions, lack of understanding, overlaps in meaning, among others. The results of the application of the process to this GT study on EdgeOps (including the different versions of the codebook and all statistics calculations of Krippendorff coefficients) are available in a public repository (see Section~\ref{sec:dataavailability}).

\subsection{Inter-coder Agreement (ICA)}\label{sec:inter-coder-agreement}

ICA is applied to a raw matter of data (typically, transcripts of interviews, answers to surveys, video data...), over which various coders highlight relevant parts (known as quotations) and label these quotations through a collection of codes (known as codebook) that represent different aspects of the reality that researchers want to understand. Additionally, codes are typically gathered into some meta-categories, called semantic domains. These semantic domains represent a facet of the reality that researchers want to understand in a broad sense. Thus, it is typical to have some semantic domains $S_1, S_2, \ldots, S_n$ and each of these domains $S_i$ contains several codes $C_{i1}, C_{i2}, \ldots, C_{in_{i}}$. This division cannot be arbitrary and must satisfy a property known as \textit{mutual exclusiveness}. This means that the semantics of the different codes within a domain must be disjoint or, in other words, it cannot be possible to assign to the same quotation two codes of the same semantic domain ($C_{ij}$ and $C_{ik}$ with $j \neq k$). To illustrate these concepts, suppose that we are analyzing the clothes of a fashion show. Then, we may have a semantic domain $S_1 = \textrm{colors}$, with inner codes $C_{11} = \textrm{black}, C_{12} = \textrm{white}, C_{13} = \textrm{red}$ and $C_{14} = \textrm{blue}$; as well as a semantic domain $S_2 = \textrm{type}$, with inner codes $C_{21} = \textrm{dress}, C_{22} = \textrm{shirt}$ and $C_{23} = \textrm{trousers}$. To each clothe (quotation), we can assign one code from $S_1$ (its main color) and one code from $S_2$ (its type), but it is not possible to apply two colors or two types to the same clothe (mutual exclusiveness).


Henceforth, at the end of the coding process, a collection of quotations has been labelled by each of the coders with one or more codes from the semantic domains according to the mutual exclusiveness rule. However, it is perfectly possible that the codings provided by the different coders do not agree, i.e., different subjects are interpreting the reality in different ways, maybe due to inconsistencies or fuzziness of the definition of the codes. To correct this issue, it is necessary to evolve the codebook by refining both codes and meanings until all coders interpret it in the same way and agree on its application. The detection of these flaws in agreement is precisely the aim of the ICA techniques. These are a collection of quantitative coefficients that allows us to measure the amount of disagreement in the different codings and to determine whether it is acceptable (so we can rely on the output of the coding process) or not (so we must refine the codebook and repeat the coding with new data). 

For this purpose, in \cite{gonzlezprieto:2020}, it was established a unified framework for measuring and evaluating the ICA based on a new interpretation of Krippendorff's $\alpha$ coefficients. Krippendorff's $\alpha$ coefficients \cite{Hayes:2007,Krippendorff:2004b,Krippendorff:2011,Krippendorff:2016} are part of a standard tool used for quantifying the agreement in content and thematic analysis due to its well-established mathematical properties and probabilistic interpretation. In our research, we shall use Krippendorff’s $\alpha$ coefficients, as described in Appendix \ref{appendix-ica}.

\subsection{Initial/Open coding}
\label{sec:open-coding}

Recall from Section \ref{sec:process} that this activity aims to discover the concepts underlying the data and instantiate in the form of codes. Thereby, at each iteration of the open coding, $n$ documents of the survey (a document is a set of answers to the survey by one of the participants) are analyzed by R1, R2, and R3, i.e., chopped into quotations that are assigned to either a previously discovered code or a new one that emerges to capture a new concept. 

The process is conducted as follows. R1 analyzes the $n$ documents, i.e., identifies quotations, creates a codebook (codes and semantic domains), and faces the coding. When R1 ends, R2 analyzes the same $n$ documents by using the codebook created by R1, i.e., analyzes previous quotations or identifies new ones and labels these quotations with a code previously proposed by R1 or new codes. If R2 adds new quotations or codes, these changes are reported in a \textit{disagreements diary}. After R2 finishes the coding process, the new codebook is delivered to R3 that repeats the process. Thereby, according to our process, the coders use the codes previously proposed or generate new ones if they think that some key information is missing. Hence, the process is flexible enough to allow the coders to add their points of view in the form of new codes, but the existence of a common codebook also increases the chances of achieving a consensus.

After an iteration ends (i.e., $n$ documents have been coded by R1, R2, and R3), the ICA is calculated. In particular, we shall use the Krippendorff's $\Cualpha$ coefficient as a quality control, with two scenarios being possible:

\begin{itemize}

\item $\Cualpha$ is below than an acceptable threshold (in our case, we fix the standard $\Cualpha < 0.8$). This evidences that there exist significant disagreements in the interpretation of the codes among the coders. In that situation, R1, R2, and R3 meet to discuss their interpretation of the codes. This \textit{review meeting} delivers the \textit{disagreements diary} and a \textit{refined codebook} in which the definitions and range of application of the codes are better delimited. With this new codebook as a basis, a new iteration starts with the next $n'$ documents of the corpus.

\item $\Cualpha$ is above or equal to the threshold ($\Cualpha \geq 0.8$). This means that there exists a consensus among the coders on the meaning of the codes. At this point, the open coding process stops and the generated codes (actually, the whole codebook) are used as input for the following activities, i.e., the selection of the core categories (Section \ref{sec:selection-core}) and the selective coding (Section \ref{sec:selective-coding}). 

\end{itemize}


Additionally, the value of the Krippendorff's $\cualpha$ coefficient is also computed per semantic domain. As explained in Section \ref{sec:inter-coder-agreement}, a low value of $\cualpha$ in a particular domain means that the coders are failing in interpreting the codes of that domain in the same way. This provides a valuable clue of the conflicting codes so that the discussion of the meaning of the codes can be focused on these codes. Thus, A small value of $\cualpha$ points out to potentially problematic codes, so that, during the review meeting, the coders can focus on the codes of these domains. Hopefully, this will lead to a more effective refinement of the codebook, which improves the ICA value of the next iteration more markedly.


The following sections describe the evolution of the agreement during the open coding activity of our GT study on EdgeOps. As we will see, after the first iteration of the coding, there was no consensus on the meaning of the codes ($\Cualpha < 0.8$). However, after refining the codebook and conducting a second iteration, the agreement improved to reach an acceptable threshold ($\Cualpha \geq 0.80$) so the initial coding was concluded.

\subsubsection*{Iteration 1}

In the first iteration of the open coding process, R1, R2, and R3 analyzed $6$ documents. R1 created a codebook with 29 codes that was subsequently refined by R2 and R3. As by-product of this process, $40$ codes were discovered and divided into $7$ semantic domains (denoted by S1, S2, ..., S7). After completing the coding process the $\Cualpha$ and $\cualpha$ ICA coefficients were computed and their values are shown in Table \ref{tab:table-open-iter1}.

\begin{table}[h]
    \begin{center}
    \small
    \caption{Values of the different Krippendorff's $\alpha$ coefficients in the iteration 1 of the open coding. In bold, the values above the acceptability threshold ($ \geq 0.80$).}
    \vspace{0.2cm}
    \begin{tabular}{|c|c|c|c|c|c|c|c|}
    \hline \multicolumn{7}{|c|}{$\cualpha$ per semantic domain} & \multirow{ 2}{*}{$\Cualpha$}\\\cline{1-7}
    S1 & S2 & S3 & S4 & S5 & S6 & S7 &  \\\hline
    \textbf{0.81} & \textbf{0.98} & 0.59 & \textbf{0.80} & \textbf{1.00} & \textbf{1.00} & \textbf{1.00} & 0.56 \\\hline
    \end{tabular}
    \label{tab:table-open-iter1}
    \end{center}
\end{table}

As we can observe from this table, the value of the global $\Cualpha$ coefficient did not reach the acceptable threshold of $0.8$. For this reason, it was necessary to conduct a review meeting for discussing disagreements and the application criteria of the different codes. The outputs of this meeting are documented in the \textit{disagreements diary} file of the \textit{open coding} folder in the public repository.

To highlight the problematic codes, we considered the $\cualpha$ coefficients computed per semantic domain. For Table \ref{tab:table-open-iter1}, we observe that domain S3 got a remarkably low value of the $\cualpha$ coefficient. A thorough look at the particular codes within S3 shows that this domain includes codes related to the functionality of the system. This is particularly a fuzzy domain in which several concepts can be confused. During the review meeting, clarifications about these codes were necessary to avoid misconceptions. After this, a new codebook was released. In this new version, memos and comments were added, and a code was removed, so $39$ codes (and $7$ semantic domains) proceeded to the second iteration of the open coding. 

\subsubsection*{Iteration 2}

R1, R2, and R3 analyzed other $6$ documents. Since the coders agreed on a common codebook in the previous iteration, we can expect a greater agreement that materializes as a higher value of ICA. As by-product of this second iteration, $8$ new codes arose leading to a new version of the codebook with $47$ codes and $7$ semantic domains. The ICA values of this second iteration are shown in Table \ref{tab:table-open-iter2}.

\begin{table}[h]
    \begin{center}
    \small
    \caption{Values of the different Krippendorff's $\alpha$ coefficients in the iteration 2 of the open coding. In bold, the values above the acceptability threshold ($ \geq 0.80$).}
    \vspace{0.2cm}
    \begin{tabular}{|c|c|c|c|c|c|c|c|}
    \hline \multicolumn{7}{|c|}{$\cualpha$ per semantic domain} & \multirow{ 2}{*}{$\Cualpha$}\\\cline{1-7}
    S1 & S2 & S3 & S4 & S5 & S6 & S7 &  \\\hline
    0.72 & \textbf{0.97} & \textbf{0.88} & \textbf{1.00} & \textbf{1.00} & \textbf{1.00} & \textbf{1.00} & \textbf{0.80} \\\hline
    \end{tabular}
    \label{tab:table-open-iter2}
    \end{center}
\end{table}

From the results of this table, we observe that, after this refinement of the codebook, the $\Cualpha$ reaches the acceptable threshold of agreement. In this way, the open coding process can stop: there exists consensus in the interpretation of the codes presented in the codebook and we can proceed with the selection of core categories and selective coding.


\subsection{Selection of core categories}
\label{sec:selection-core}

\begin{figure}[h]
\centering
\includegraphics [width=9cm]{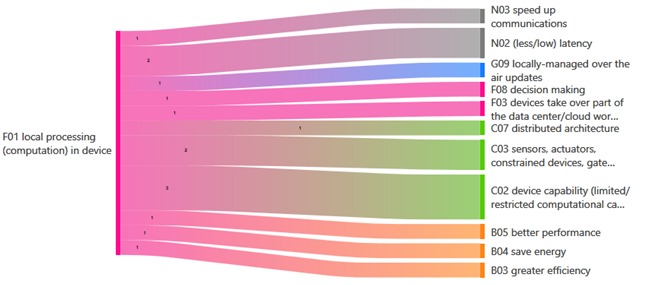}
\caption{Illustrative example of density analysis}
\label{fig:density}
\end{figure}

In this activity, R1 and R2 selected the core categories, i.e., the most relevant codes from the $47$ codes obtained in the open coding. To that end, we focused on the groundedness of the codes and semantic domains (i.e., the number of quotations coded by a code) and the density of the codes and semantic domains (i.e., the number of relationships between codes, that means, the co-occurrence of codes in the same quotation). The detailed analysis is documented in the \textit{selection of core categories} file of the \textit{selection of core categories} folder in the public repository. Figure~\ref{fig:density} shows an example of the multiple tables and graphics obtained from Atlas.ti that were analyzed during this activity. In this figure, the code ''F01 local processing´´ is related to 11 codes 15 times. As a result of the analysis, four semantic (S1, S2, S3 and S6) domains and $29$ codes were selected for the next activity. This codebook is available in the \textit{selection of core categories - codebook} file of the public repository.
 
\subsection{Selective coding}
\label{sec:selective-coding}

Recall from Section \ref{sec:process} that this is an inductive-deductive process in which new data are labelled with the codes of selected categories (semantic domains). Three coders (R1, R2, and R3) were involved again in this activity. The coders only focused on the core categories, but the number and definition of their inner codes were modified according to the analysis of new data.


After an iteration ends (i.e., $n$ documents have been coded by R1, R2, and R3), the ICA is calculated. If the value of $\Cualpha$ is below the threshold of acceptability of $0.8$, the coders meet as in Section \ref{sec:open-coding} to refine the codebook. After polishing this new version of the codebook, a new iteration of selective coding is conducted to check whether they reach an acceptable agreement.

However, even in the case that $\Cualpha$ passes the acceptability threshold, it may happen that some extra iterations of the coding process are needed. Indeed, to proceed with the following activity (sorting), it is mandatory that the new data analyzed do not introduce new information to the theory (the so-called theoretical saturation). For this reason, even if $\Cualpha \geq 0.8$, the coders must have a meeting to discuss whether theoretical saturation has been reached. If they decide that the saturation is not fulfilled yet, an extra iteration of selective coding must be conducted. After completing this new iteration, both the ICA (via $\Cualpha$) and the theoretical saturation are analyzed. Only when both the ICA and the saturation are satisfactory, the GT process can proceed to the next activity.

In the GT study here described, only one iteration was needed to fulfil both the ICA and the saturation criteria.

\subsubsection*{Iteration 1}

In this iteration, R1, R2, and R3 analyzed 6 documents using S1, S2, S3, and S6, which encompass a total of $29$ codes. 
After coding, $9$ codes were added to the codebook accounting for a total of $38$ core codes. This codebook is available in the \textit{codebook} file of the \textit{selective coding} folder in the public repository. The results of the ICA coefficients obtained after coding are shown in Table \ref{tab:table-selective}.

\begin{table}[h]
    \begin{center}
    \small
    \caption{Values of the different Krippendorff's $\alpha$ coefficients in Iteration 1 of the selective coding phase. In bold, the values above the acceptability threshold ($ \geq 0.80$).}
    \vspace{0.2cm}
    \begin{tabular}{|c|c|c|c|c|}
    \hline \multicolumn{4}{|c|}{$\cualpha$ per semantic domain} & \multirow{ 2}{*}{$\Cualpha$}\\\cline{1-4}
    S1 & S2 & S3 & S6 &  \\\hline
    \textbf{1.00} & \textbf{0.95} & \textbf{0.87} & \textbf{1.00} & \textbf{0.80} \\\hline
    \end{tabular}
    \label{tab:table-selective}
    \end{center}
\end{table}

As we can observe from this table, the value of $\Cualpha$ did reach the acceptable threshold of reliability of $0.8$. This evidences that there exists a consensus among the coders on the meaning and limits of the codes within the core categories. Additionally, the coders also agreed that adding new data did not lead to new information, so the theoretical saturation had been reached. Therefore, since after this first iteration, the value of $\Cualpha$ was compelling and the coders agreed that the theoretical saturation had been reached, the GT process could proceed to the next activity.

At this point, the proposed GT process coincides with the existing approaches in the literature: a sorting procedure followed by theoretical coding during which a theory emerges. Since the focus of this work is to improve the rigor and consensus of the elicited codes during the open and selective coding procedures, for the sake of simplicity, we skip these subsequent standard GT phases. 

\section{Threats to Validity and Limitations}
\label{sec:threats}

The meta-science standard (see Empirical Standards for
Software Engineering Research \cite{ralph2021}) guided us to analyze the issues and challenges of the GT method when various raters are involved in coding procedures and formalize a process to improve collaborative team science and consortia. To that end, we previously conducted a systematic literature review (see Section~\ref{sec:slr}), and subsequently applied the process to a GT study (see Section~\ref{sec:casestudy}). This section describes the threats to validity and limitations in both the SLR and the application case we addressed.  

There are some techniques to mitigate sampling and publication bias in SLR that we did not address, such as backward and forward snowballing searches, searching on indexes (e.g.\ Google Scholar) in addition to formal databases, and  searching for relevant dissertations or preprint servers (e.g.\ arXiv). Nevertheless, on the basis of the obtained results, we consider that the narrative synthesis and empirical evidence from the selected 49 primary studies (those which fit the inclusion and exclusion criteria out of 168 unduplicated scientific papers) were enough to answer RQ1 and RQ2. A larger sample would not have provided new findings, but would have strengthened the evidence. 

Quantitative quality criteria such as internal validity and construct validity do not apply, as this is not the kind of SLR that conducts meta-analysis to aggregate data for causal relationships between constructs. However, we do provide \textit{replication package} including search terms and results, selection process results, examples of coding, and complete synthesis results. The selection process (i.e., the application of inclusion and exclusion criteria) was sufficiently rigorous for the systematic review goals as two researchers participated in a dual selection process (as described in Section~\ref{subsec:selection}) and IRA (specifically, Krippendorff’s $\alpha$ binary)  was iteratively analyzed (as described in Section~\ref{subsec:selectionresults}) to improve inclusion and exclusion criteria.

Conclusion validity concerns the relationship between treatment and outcomes \cite{wohlin:2012}, for example, how different researchers might have addressed data extraction and data synthesis differently. In our case, two researchers with different backgrounds independently extracted the data from primary papers without duplicity, as we considered that duality was not necessary as the data to be extracted are totally objective. We provide some coding examples using Atlas.ti available in the replication package. Additionally, we have extensively used quotations to establish credibility in the qualitative sense of chain-of-evidence (see Section~\ref{subsec:datasynthesis}).  

Finally, regarding the process for IRR/IRA in GT and its application to a GT study in the domain of \textit{EdgeOps in industry}, the main concern is about the external validity and generalizability, which typically does not apply to case studies in which the effort to demonstrate feasibility is enormous, as it requires the execution of multiple cases (i.e., multiple GT studies) from data collection to theory generation. Thus, we can only assert that the application of this process to a GT study seems to support its feasibility. In the absence of further confirmation, this would represent the first step of a de facto standard to be applied to those GT studies that require IRR/IRA. 

\section{Related Work}
\label{sec:relatedwork}

Sharing our objectives, several authors have tried to systematize the role of multiple coders and agreement measures in the qualitative research paradigm, ranging from phenomenology \cite{marques:2005}, content analysis \cite{nili:2020} to constant comparative methods \cite{olson:2016}. 

Olson et al.\ \cite{olson:2016} propose a methodology for constant comparative method, IRR/IRA and multiple researchers with agreement measured through Fleiss' $\kappa$ coefficient. In this work, the authors ponder upon the role of IRR/IRA not as a verification tool but as a \textit{solidification} tool, a concept borrowed form the application of IRR/IRA into the constructivist phenomenologist paradigm, following \cite{marques:2005}. They also address the inclusion of a positivist term like ''reliability'' in the qualitative paradigm and focus on the use of IRR/IRA for providing transparency by means of a ''clear protocol, codebook, and database''. The method is applied to the constant comparative method, underlying a \textit{hard} positivist epistemology. It is remarkable that the authors report that they felt so constrained with the use of IRR/IRA during codification that it led to loss of meaning. The search for a \textit{good} value for IRR/IRA distorted the purpose of codification to the extent of being more concerned with coincidence with other researchers than with meaningful codification. Thus, they shifted the interpretation of Fleiss' $\kappa$ from a quantitative verification tool to a tool to ``guide collaboration and identify nuances in the data brought to light by our prior experience, knowledge, and perspectives''. 

In the aforementioned work, the authors outline the 10-steps method (with possible iterations) that they followed during the experience: 
\begin{enumerate}
    \item Each researcher performed open-coding of Week 1 logs. 
    \item Collaborated to unify codes. 
    \item Each researcher re-coded Week 1 logs using unified codes. 
    \item Calculated ICR (IRR/IRA). 
    \item Collaborated to discuss each code and identify areas lacking agreement. 
    \item Repeated the above process for each week of logs, producing a unified codebook applicable to all logs. 
    \item Re-coded all logs, producing themes. 
    \item Selected themes for further analysis. 
    \item Conducted co-occurrence analysis. 
    \item Constructed an exploratory model – the findings of the study 
\end{enumerate}

As we can observe from this method, there is an open coding phase performed by all coders to create a first version of the codebook which is not subject to ICR calculation. Then, the same data is re-codified and IRR/IRA is calculated. A threshold for agreement is not sought, but it is used as a tool for unveiling disagreement areas and possible behavior patterns of coders (step 5). These five steps are repeated for all data samples, which means that before entering into the next phase of coding (selective or thematic), the data is passed over and codified several times.

The inclusion of quantitative techniques into qualitative research and the need to follow clear guidelines and a sound methodology in the discipline of Information Systems is thoroughly revised in \cite{venkatesh:2013}. There, they developed a meta-tool guiding authors to combine both methods, aiming at high level epistemological issues that researchers should approach when combing methods. Closer to our interest is the work of Nili et al.\ \cite{nili:2020}, which focuses on the practical issues of applying IRR/IRA to qualitative methods (also circumscribed to Information Systems discipline). In this work, the authors provide guidelines to decide on the most suitable statistical instrument for IRR/IRA and a 5-step approach to perform IRA/IRR in qualitative studies, namely: 

\begin{enumerate}
  \item Selecting an ICR method 
  \item	Developing a coding scheme
  \item Selecting and training independent coders, 
  \item	Calculating the ICR coefficient (which may lead to continuing the training session and iteratively coding the entire dataset), 
  \item	Reporting the process of evaluating ICR along with the result. 
\end{enumerate}

With a possible cycle of iterations, like Olson, from step 4 to step 2, both works coincide in excluding IRR/IRA calculation from the first phase of open-coding phase. Open coding of Step 2 is performed by only one coder and the codebook is said to be constructed both inductively (from raw data) or derived from previous literature. Therefore, in this methodology IRR/IRA seems to play a secondary role at first (or even no role), and it is used as a posteriori checking/verification. 

\section{Conclusion}
\label{sec:conclusion}

Qualitative research, and GT as one instance, is often tarnished by epistemological debates like the validity and reliability of obtained knowledge. When applied to computer science (CS), the epistemological position is usually not clear. However, it is not infrequent to apply quantitative instruments in qualitative research as a possibly a way to confer validity and methodological strength to the researches carried out in the qualitative paradigm. We have focused on the use of quantitative instruments like different IRR/IRA techniques applied to GT studies. As shown in Section~\ref{sec:slr}, GT-driven research in CS usually presents some deficiencies when dealing with the epistemological and methodological issues that support the validity and reliability of their outcomes: self-allegedly GT studies do not clarify which GT school/trend they adhere to, thus, using GT terminology confusingly and reducing GT methodology to the mere use of coding procedures. Besides, IRR/IRA instruments are sometimes poorly used, like confusing the concepts of reliability and agreement, and above all, using these statistical instruments with no further purpose in the study.

Convinced as we are of the utility and essential role in the science of qualitative research and aware of the validity and quality issues of the obtained results, we have formalized a process to integrate IRR/IRA into qualitative GT-based research that allows researchers to rigorously use these statistical techniques for measuring reliability and agreement during the coding process, and thus, fostering consensus and reflection. We do not consider the notion of reliability as trying to establish a single reality, but rather as an approach for developing a shared understanding that can also establish consistency among coders. Our method is independent from, and should fit, different families of GT theory. It is targeted to those who decide to validate consensus and shared understanding in teamwork during coding processes. The process was validated with a case of study (delimited in scope and extension) to prove its feasibility. It is a delimited case study in extension but not in depth, focusing on open and selective coding phases. 
Finally, there is not a definite and correct way to handle validity and reliability in qualitative research. Our intention is to define the first steps towards the definition of a \textit{de facto} standard to be applied to those GT studies that would benefit from the application of IRR/IRA instruments.

\section*{Data Availability}
\label{sec:dataavailability}
Link to supplementary materials in a long-term archive: \url{https://doi.org/10.5281/zenodo.5034244}.
Data of the GT study on EdgeOps were collected from a open-ended questionnaire available in \url{https://es.surveymonkey.com/r/PMWD7ZM}.

\section*{Acknowledgments}
The authors would like to thank Paul Ralph for his valuable review and suggestions. The forth named author acknowledges the hospitality of the Department of Mathematics at Universidad Aut\'onoma de Madrid, where part of this work was completed.

\appendix
\section{Krippendorff's $\alpha$ for ICA}
\label{appendix-ica}

\begin{figure*}[h!]
\centering
\includegraphics [width=14cm]{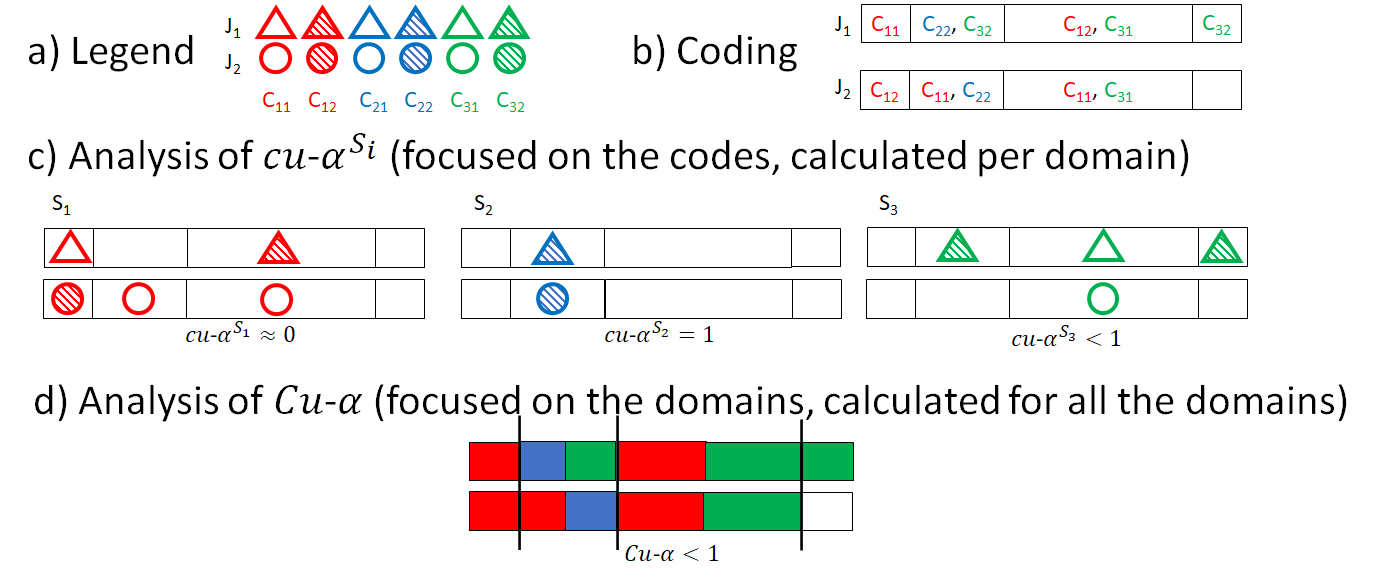}
\caption{Illustrative example of the Krippendorff's $\alpha$ coefficients}
\label{fig:7}
\end{figure*}

This appendix describes the following two versions of Krippendorff’s $\alpha$ coefficients:

\begin{itemize}

\item The coefficient $\cualpha$: This coefficient is computed on a specific semantic domain $S$. It indicates the degree of agreement with which coders identify codes within $S$.

\item The coefficient $\Cualpha$: This coefficient measures the degree of agreement in the decision to apply different semantic domains, independent of the chosen code.
\end{itemize}

For the convenience of the reader, we provide a running example of the use of these coefficients. This case of use has been extracted from \cite{diaz2020} (see also \cite{perez2021devops}. Figure~\ref{fig:7} shows an illustrative example of the use of these coefficients. Let three semantic domains and their respective codes be as follows:
$$
	S_1 = \left\{C_{11}, C_{12}\right\}, \quad S_2 = \left\{C_{21}, C_{22}\right\}, \quad S_3 = \left\{C_{31}, C_{32}\right\}.
$$

Coder 1 and Coder 2 assign codes to four quotations, as shown in Figure~\ref{fig:7}(a), such that the first quotation is assigned ${C_{11}}$ by Coder 1 and ${C_{12}}$ is assigned by Coder 2. We create a graphical metaphor so that each coder, each semantic domain, and each code are represented as shown in Figure~\ref{fig:7}(b). Each coder is represented by a shape, such that Coder 1 is represented by triangles and Coder 2 is represented by circles. Each domain is represented by a color: $S_1$ is red, $S_2$ is blue, and $S_3$ is green. Each code within the same semantic domain is represented as a fill, where ${C_{i1}}$ codes are represented by a solid fill and ${C_{i2}}$ codes are represented by dashed fills.

The coefficient $\cualpha$ is calculated per domain (i.e., $S_1$ red, $S_2$ blue, $S_3$ green), but it measures the agreement attained when applying the codes of that domain. In other words, given a domain $S_i$, this coefficient analyzes whether the coders assigned the same codes of $S_i$ (i.e., the same type of fill) to the quotations or not. In this way,  Figure~\ref{fig:7}(c) only focuses on the fills applied to each quotation. In particular, it is shown that $\cualpha=1$ for $S_2$, since both coders assigned the same code to the second quotation, but no code from this domain was assigned to the rest of the quotations, i.e., total agreement. Additionally, it is shown that $\cualpha<1$ for $S_3$, as the coders assigned the same code of  $S_3$ to the third quotation 3, but they did not assign the same codes of  $S_3$ to the rest of the quotations. Finally, it is shown that the $\cualpha$ coefficient for $S_1$ is very small (near zero) since the coders achieved no agreement on the chosen codes (the exact value of $\cualpha$ will depend on the expected disagreement, which depends on the marginal frequencies of each code).

On the other hand, the coefficient $\Cualpha$ analyzes opera
opera  all domains as a whole, but it does not take into account the codes within each domain. In this way, in Figure~\ref{fig:7}(d), we color each segment with the colors corresponding to the applied semantic domain (regardless of the particular code used). From these chromatic representations, $\Cualpha$ measures the agreement in applying these colors globally among the coders. In particular, note that $\Cualpha<1$, as both coders assigned the same domain $S_1$ to the first quotations, and  they assigned domains $S_1$ and $S_3$ to the third quotation, but they did not assign the same domains in the second and fourth quotations.

The larger the $\alpha$ coefficients are, the better the observed  agreement. Typically, the $\alpha$ coefficients lie in the range of $0 \leq \alpha \leq 1$. A common rule-of-thumb in the literature \cite{Krippendorff:2018} is that $\alpha \geq 0.667$ is the minimal threshold required for drawing conclusions from the data. For $\alpha \geq 0.80$, we can consider that there exists statistical evidence of reliability in the coding. A thorough explanation of the use of these coefficients and their interpretation can be found in \cite{gonzlezprieto:2020}.

\bibliographystyle{spmpsci}      
\bibliography{references}   

\begin{thebibliography}{10}
\providecommand{\url}[1]{{#1}}
\providecommand{\urlprefix}{URL }
\expandafter\ifx\csname urlstyle\endcsname\relax
  \providecommand{\doi}[1]{DOI~\discretionary{}{}{}#1}\else
  \providecommand{\doi}{DOI~\discretionary{}{}{}\begingroup
  \urlstyle{rm}\Url}\fi

\bibitem{armstrong:1997}
Armstrong, D., Gosling, A., Weinman, J., Marteau, T.: The place of inter-rater
  reliability in qualitative research: An empirical study.
\newblock Sociology \textbf{31}(3), 597--606 (1997).
\newblock \doi{10.1177/0038038597031003015}.
\newblock \urlprefix\url{https://doi.org/10.1177/0038038597031003015}

\bibitem{braun:2013}
Braun, V., Clarke, V.: Successful qualitative research.
\newblock Sage (2013)

\bibitem{campbell:2013}
Campbell, J.L., Quincy, C., Osserman, J., Pedersen, O.K.: Coding in-depth
  semistructured interviews: Problems of unitization and intercoder reliability
  and agreement.
\newblock Sociological Methods \& Research \textbf{42}(3), 294--320 (2013).
\newblock \doi{10.1177/0049124113500475}.
\newblock \urlprefix\url{https://doi.org/10.1177/0049124113500475}

\bibitem{charmaz:2014}
Charmaz, K.: Constructing Grounded Theory.
\newblock Sage 2nd Ed. (2014)

\bibitem{Charmaz:2020}
Charmaz, K., Thornberg, R.: The pursuit of quality in grounded theory.
\newblock Qualitative Research in Psychology \textbf{0}(0), 1--23 (2020).
\newblock \doi{10.1080/14780887.2020.1780357}

\bibitem{cohen1960coefficient}
Cohen, J.: A coefficient of agreement for nominal scales.
\newblock Educational and psychological measurement \textbf{20}(1), 37--46
  (1960)

\bibitem{creswell:2017}
Creswell, J.W., Creswell, J.D.: Research design: Qualitative, quantitative, and
  mixed methods approaches.
\newblock Sage publications (2017)

\bibitem{cruzes:2011}
Cruzes, D.S., Dyba, T.: Recommended steps for thematic synthesis in software
  engineering.
\newblock In: 2011 international symposium on empirical software engineering
  and measurement, pp. 275--284. IEEE (2011)

\bibitem{diaz2020}
D{\'\i}az, J., L{\'o}pez-Fern{\'a}ndez, D., P{\'e}rez, J., Gonz{\'a}lez-Prieto,
  {\'A}.: Why are many businesses instilling a devops culture into their
  organization?
\newblock Empirical Software Engineering \textbf{26}(2), 1--50 (2021)

\bibitem{dube2003rigor}
Dub{\'e}, L., Par{\'e}, G.: Rigor in information systems positivist case
  research: current practices, trends, and recommendations.
\newblock MIS quarterly pp. 597--636 (2003)

\bibitem{erickson:1998}
Erickson, K., Stull, D.: Doing team ethnography: Warnings and advice.
\newblock Thousand Oaks, CA: Sage (1998)

\bibitem{Gibbs:2007}
Gibbs, G.R.: Analyzing qualitative data. Qualitative Research Kit.
\newblock Sage publications (2007)

\bibitem{gisev2013interrater}
Gisev, N., Bell, J.S., Chen, T.F.: Interrater agreement and interrater
  reliability: key concepts, approaches, and applications.
\newblock Research in Social and Administrative Pharmacy \textbf{9}(3),
  330--338 (2013)

\bibitem{glaser:1967}
Glaser, B., Strauss, A.L.: The Discovery of Grounded Theory: Strategies for
  Qualitative Research.
\newblock Aldine de Gryter, New York (1967)

\bibitem{gonzlezprieto:2020}
Ángel González-Prieto, Perez, J., Díaz, J., López-Fernández, D.:
  Inter-coder agreement for improving reliability in software engineering
  qualitative research (2020).
\newblock Https://arxiv.org/abs/2008.00977

\bibitem{guest:2008}
Guest, G., MacQueen, K.M.: Handbook for team-based qualitative research.
\newblock Lanham, MD: AltaMira Press (2008)

\bibitem{hammer:2014}
Hammer, D., Berland, L.K.: Confusing claims for data: A critique of common
  practices for presenting qualitative research on learning.
\newblock Journal of the Learning Sciences \textbf{23}(1), 37--46 (2014).
\newblock \doi{10.1080/10508406.2013.802652}.
\newblock \urlprefix\url{https://doi.org/10.1080/10508406.2013.802652}

\bibitem{Hayes:2007}
Hayes, A.F., Krippendorff, K.: Answering the call for a standard reliability
  measure for coding data.
\newblock Communication Methods and Measures \textbf{1}(1), 77--89 (2007).
\newblock \doi{10.1080/19312450709336664}

\bibitem{hoda2021}
Hoda, R.: Decoding grounded theory for software engineering.
\newblock In: 2021 IEEE/ACM 43rd International Conference on Software
  Engineering: Companion Proceedings (ICSE-Companion), pp. 326--327 (2021).
\newblock \doi{10.1109/ICSE-Companion52605.2021.00139}

\bibitem{hoda2012}
Hoda, R., Noble, J., Marshall, S.: Developing a grounded theory to explain the
  practices of self-organizing agile teams.
\newblock Empirical Software Engineering \textbf{17}(6), 609--639 (2012).
\newblock \doi{10.1007/s10664-011-9161-0}.
\newblock \urlprefix\url{https://doi.org/10.1007/s10664-011-9161-0}

\bibitem{Kenny:2015}
Kenny, M., Fourie, R.: Contrasting classic, straussian, and constructivist
  grounded theory: Methodological and philosophical conflicts.
\newblock The Qualitative Report \textbf{20}(8), 1270--1289 (2015).
\newblock \doi{https://doi.org/10.46743/2160-3715/2015.2251}

\bibitem{kitchenham2007guidelines}
Kitchenham, B.A., Charters, S.: Guidelines for performing systematic literature
  reviews in software engineering.
\newblock Tech. Rep. EBSE 2007-001, Keele University and Durham University
  Joint Report (2007).
\newblock
  \urlprefix\url{https://www.elsevier.com/__data/promis_misc/525444systematicreviewsguide.pdf}

\bibitem{Krippendorff:2004b}
Krippendorff, K.: Reliability in content analysis: Some common misconceptions
  and recommendations.
\newblock Human Communication Research \textbf{30}(3), 411--433 (2004).
\newblock \doi{10.1111/j.1468-2958.2004.tb00738}.
\newblock \urlprefix\url{https://doi.org/10.1111/j.1468-2958.2004.tb00738}

\bibitem{Krippendorff:2011}
Krippendorff, K.: Computing krippendorff's alpha-reliability (2011)

\bibitem{Krippendorff:2018}
Krippendorff, K.: Content analysis: An introduction to its methodology, 4th
  edn.
\newblock Sage publications (2018)

\bibitem{Krippendorff:2016}
Krippendorff, K., Mathet, Y., Bouvry, S., Widlöcher, A.: {On the reliability
  of unitizing textual continua: Further developments}.
\newblock Quality \& Quantity: International Journal of Methodology
  \textbf{50}(6), 2347--2364 (2016).
\newblock \doi{10.1007/s11135-015-0266-1}.
\newblock
  \urlprefix\url{https://ideas.repec.org/a/spr/qualqt/v50y2016i6d10.1007_s11135-015-0266-1.html}

\bibitem{Lincoln:1985}
Lincoln, Y., Guba, E.: Naturalistic inquiry.
\newblock Beverly Hills, CA: Sage (1985)

\bibitem{MacPhail:2016}
MacPhail, C., Khoza, N., Abler, L., Ranganathan, M.: Process guidelines for
  establishing intercoder reliability in qualitative studies.
\newblock Qualitative Research \textbf{16}(2), 198--212 (2016).
\newblock \doi{10.1177/1468794115577012}.
\newblock \urlprefix\url{https://doi.org/10.1177/1468794115577012}

\bibitem{marques:2005}
Marques, J., Mccall, C.: The application of interrater reliability as a
  solidification instrument in a phenomenological study.
\newblock The Qualitative Report \textbf{10} (2005).
\newblock \doi{10.46743/2160-3715/2005.1837}

\bibitem{mcdonald:2019}
McDonald, N., Schoenebeck, S., Forte, A.: Reliability and inter-rater
  reliability in qualitative research: Norms and guidelines for cscw and hci
  practice.
\newblock Proc. ACM Hum.-Comput. Interact. \textbf{3}(CSCW) (2019).
\newblock \doi{10.1145/3359174}.
\newblock \urlprefix\url{https://doi.org/10.1145/3359174}

\bibitem{nili:2017}
Nili, A., Tate, M., Barros, A.: A critical analysis of inter-coder reliability
  methods in information systems research.
\newblock In: Proceedings of the 28th Australasian Conference on Information
  Systems, pp. 1--11. University of Tasmania (2017)

\bibitem{nili:2020}
Nili, A., Tate, M., Barros, A., Johnstone, D.: An approach for selecting and
  using a method of inter-coder reliability in information management research.
\newblock International Journal of Information Management \textbf{54}, 102154
  (2020).
\newblock \doi{https://doi.org/10.1016/j.ijinfomgt.2020.102154}

\bibitem{olson:2016}
Olson, J., McAllister, C., Grinnell, L., Walters, K., Appunn, F.: Applying
  constant comparative method with multiple investigators and inter-coder
  reliability.
\newblock The Qualitative Report \textbf{21}, 26--42 (2016).
\newblock \doi{10.46743/2160-3715/2016.2447}

\bibitem{oconnor:2020}
O’Connor, C., Joffe, H.: Intercoder reliability in qualitative research:
  Debates and practical guidelines.
\newblock International Journal of Qualitative Methods \textbf{19},
  1609406919899220 (2020).
\newblock \doi{10.1177/1609406919899220}.
\newblock \urlprefix\url{https://doi.org/10.1177/1609406919899220}

\bibitem{perez2021devops}
Perez, J., Gonzalez-Prieto, A., Diaz, J., Lopez-Fernandez, D., Garcia-Martin,
  J., Yague, A.: Devops research-based teaching using qualitative research and
  inter-coder agreement.
\newblock IEEE Transactions on Software Engineering  (2021)

\bibitem{PEREZ2020110657}
Pérez, J., Díaz, J., Garcia-Martin, J., Tabuenca, B.: Systematic literature
  reviews in software engineering—enhancement of the study selection process
  using cohen’s kappa statistic.
\newblock Journal of Systems and Software \textbf{168}, 110657 (2020).
\newblock \doi{https://doi.org/10.1016/j.jss.2020.110657}.
\newblock
  \urlprefix\url{https://www.sciencedirect.com/science/article/pii/S0164121220301217}

\bibitem{ralph2021}
Ralph, P.e.: Empirical standards for software engineering research (2021).
\newblock \urlprefix\url{arXiv:2010.03525v2 [cs.SE]}

\bibitem{Saldana2012}
Salda{\~n}a, J.: The Coding Manual for Qualitative Researchers.
\newblock Sage publications (2012)

\bibitem{SALLEH:2018}
Salleh, N., Hoda, R., Su, M.T., Kanij, T., Grundy, J.: Recruitment, engagement
  and feedback in empirical software engineering studies in industrial
  contexts.
\newblock Information and Software Technology \textbf{98}, 161 -- 172 (2018).
\newblock \doi{https://doi.org/10.1016/j.infsof.2017.12.001}.
\newblock
  \urlprefix\url{http://www.sciencedirect.com/science/article/pii/S0950584917303786}

\bibitem{STOL:2016}
Stol, K.J., Ralph, P., Fitzgerald, B.: Grounded theory in software engineering
  research: A critical review and guidelines.
\newblock In: Proceedings of the 38th International Conference on Software
  Engineering, ICSE ’16, p. 120–131. Association for Computing Machinery,
  New York, NY, USA (2016).
\newblock \doi{10.1145/2884781.2884833}.
\newblock \urlprefix\url{https://doi.org/10.1145/2884781.2884833}

\bibitem{strauss:1990}
Strauss, A., Corbin, J.: Basics of Qualitative Research: Grounded Theory
  Procedures and Techniques.
\newblock SAGE Publication, London (1990)

\bibitem{venkatesh:2013}
Venkatesh, V., Brown, S.A., Bala, H.: Bridging the qualitative-quantitative
  divide: Guidelines for conducting mixed methods research in information
  systems.
\newblock MIS Quarterly \textbf{37}(1), 21--54 (2013)

\bibitem{weston:2001}
Weston, C., Gandell, T., Beauchamp, J., McAlpine, L., Wiseman, C., Beauchamp,
  C.: Analyzing interview data: The development and evolution of a coding
  system.
\newblock Qualitative Sociology \textbf{24}(3), 381–400 (2001).
\newblock \doi{10.1023/10.1023\%2FA\%3A10106909082000}.
\newblock \urlprefix\url{https://doi.org/10.1023\%2FA\%3A1010690908200}

\bibitem{wohlin:2012}
Wohlin, C., Runeson, P., H{\"o}st, M., Ohlsson, M.C., Regnell, B., Wessl{\'e}n,
  A.: Experimentation in software engineering.
\newblock Springer Science \& Business Media (2012)

\bibitem{wu:2016}
Wu S., W., C., D., Fraser, M.W.: Author guidelines for manuscripts reporting on
  qualitative research.
\newblock Journal of the Society for Social Work and Research \textbf{7},
  405–425 (2016).
\newblock \doi{10.1086/685816}.
\newblock \urlprefix\url{https://doi.org/10.1086/6858161}

\bibitem{yin2017Case}
Yin, R.: Case Study Research and Applications: Design and Methods, 6th edn.
\newblock Sage (2017)

\end{thebibliography}

\end{document}